\documentclass[a4paper,onecolumn,11pt,accepted=2025-11-04]{quantumarticle}
\pdfoutput=1

\usepackage{graphicx}
\usepackage[normalem]{ulem}
\usepackage{color,bm}
\usepackage[caption=false]{subfig}
\usepackage[numbers, sort&compress]{natbib}
\usepackage{latexsym}
\usepackage{array,multirow}
\usepackage{hyperref}
\usepackage[utf8]{inputenc}
\usepackage{amsmath}

\definecolor{xred}{rgb}{1,0,0}

\definecolor{xblue}{rgb}{0,0,1}
\def  \Blue#1{\textcolor{xblue}{#1}}
\definecolor{xgreen}{rgb}{0,1,0}
\def  \Green#1{\textcolor{xgreen}{#1}}

\begin{document}

\title{Multiple quantum exceptional, diabolical, and hybrid points in multimode bosonic systems: II.
Nonconventional $\mathcal{PT} $-symmetric dynamics and
unidirectional coupling}

\author{Jan Pe\v{r}ina Jr.}
\email{jan.perina.jr@upol.cz} \affiliation{Joint Laboratory of
Optics, Faculty of Science, Palack\'{y} University, Czech
Republic, 17. listopadu 12, 771~46 Olomouc, Czech Republic}

\author{Kishore Thapliyal}
\email{kishore.thapliyal@upol.cz}
\orcid{0000-0002-4477-6041}
\affiliation{Joint Laboratory of
Optics, Faculty of Science, Palack\'{y} University, Czech
Republic, 17. listopadu 12, 771~46 Olomouc, Czech Republic}

\author{Grzegorz Chimczak}
\affiliation{Institute of Spintronics and Quantum Information,
Faculty of Physics, Adam Mickiewicz University, 61-614 Pozna\'{n},
Poland}

\author{Anna Kowalewska-Kud\l{}aszyk}
\affiliation{Institute of Spintronics and Quantum Information,
Faculty of Physics, Adam Mickiewicz University, 61-614 Pozna\'{n},
Poland}

\author{Adam Miranowicz}
\affiliation{Institute of Spintronics and Quantum Information,
Faculty of Physics, Adam Mickiewicz University, 61-614 Pozna\'{n},
Poland}
\maketitle

\begin{abstract}
{ We analyze the existence and degeneracies of quantum
exceptional, diabolical, and hybrid points in simple bosonic
systems - comprising up to six modes with damping and/or
amplification - under two complementary scenarios to those of
Ref.~\cite{Thapliyal2025}: (i) nonconventional PT-symmetric
dynamics confined to a subspace of the full Liouville space, and
(ii) systems featuring unidirectional coupling.} The system
dynamics described by quadratic non-Hermitian Hamiltonians is
governed by the Heisenberg-Langevin equations. Conditions for the
observation of inherited quantum hybrid points with up to
sixth-order exceptional and second-order diabolical degeneracies
are revealed, though relevant only for short-time dynamics. This
raises the question of whether higher-order inherited
singularities exist in bosonic systems under general conditions.
Nevertheless, for short times, unidirectional coupling of various
types enables the concatenation of simple bosonic systems with
second- and third-order exceptional degeneracies such that
arbitrarily high exceptional degeneracies are reached. Methods for
numerical identifying the quantum exceptional and hybrid points
together with their degeneracies are addressed. Following
Ref.~\cite{Thapliyal2025} rich dynamics of second-order
field-operator moments is analyzed from the point of view of the
presence of exceptional and diabolical points and their
degeneracies.
\end{abstract}

\maketitle

\section{Introduction}

Non-Hermitian bosonic parity-time ($ \mathcal{PT} $) -symmetric
systems exhibit a range of remarkable phenomena at and near
quantum exceptional and hybrid points (QEPs and QHPs,
respectively). Several
studies~\cite{Chen2017,Liu2016,Feng2017,Hodaei2017,El-Ganainy2019,Parto2021}
have shown that these singularities can enhance measurement
precision beyond the classical limit and amplify effective system
nonlinearities, leading to the generation of highly nonclassical
and entangled states with tailored properties
\cite{He2015,Vashahri2017,PerinaJr2019b,PerinaJr2019c}. {We note
that several studies (e.g.,~\cite{Langbein2018,Mortensen2018,
Zhang2019,Chen2019,Naikoo2023,Loughlin2024}) have demonstrated
that in the linear regimes of quantum systems the measurement
sensitivity boost may not occur. At present, investigations of
QEPs in nonlinear quantum systems from the point of view of the
measurement precision proceed to demonstrate the advantage of $
\mathcal{PT} $-symmetric dynamics in connection with the system
nonlinearity that allows on its own to beat the classical limit.}
However, the intrinsic noise accompanying damping and
amplification in $ \mathcal{PT} $-symmetric quantum systems must
also be taken into account, as it typically degrades quantumness
over extended timescales~\cite{Scheel2018,PerinaJr2023}. Despite
this, highly squeezed and sub-Poissonian states remain achievable.
Under various configurations of passive and active $ \mathcal{PT}
$-symmetric bosonic systems, one can also generate entangled
states exhibiting asymmetric steering and Bell
nonlocality~\cite{PerinaJr2025}. Moreover, these systems --- and
their field-operator moments (FOMs) \cite{Arkhipov2023} --- can
simulate complex many-body dynamics, including boundary
effects~\cite{Arkhipov2023,Arkhipov2021a}. On the applied side,
optical switching via encirclement of a QHP has been
demonstrated~\cite{Arkhipov2023a}, and unidirectional light
propagation --- along with invisibility cloaking --- has been
realized in such
platforms~\cite{Peng2014,Chang2014,Lin2011,Regen2012}. For a
comprehensive overview of the rich physics around the spectral
singularities of $ \mathcal{PT} $-symmetric systems, see the
reviews in~\cite{Ozdemir2019,Miri2019}.

The strength of these effects depends in many cases on the order
of the degeneracy of exceptional points (EPs): The higher-order
the degeneracy, the more enhanced the processes become. This led
us in Part I of Ref.~\cite{Thapliyal2025} to the analysis of QEPs
and QHPs of $ \mathcal{PT} $-symmetric bosonic systems with up to
five modes considering different configurations {under very
general conditions.} However, this analysis revealed only the
systems with QHPs with the second- and third-order exceptional
degeneracies (EDs) and second-order diabolical degeneracies (DDs),
despite the fact that the bosonic systems with five modes were
considered with the promise of observation of QHPs with the
fifth-order ED.

For this reason, we consider more general bosonic system compared
to those analyzed in~\cite{Thapliyal2025} to seek for the
observation of higher-order EDs for inherited QEPs and QHPs.
First, we weaken our requirements for the observation of $
\mathcal{PT} $-symmetric dynamics by considering only subspace(s)
of the whole Liouville space of the statistical operators. We note
here that, owing to the linearity of quantum mechanics, we can
equivalently describe the system dynamics
\cite{PerinaJr1995,Vogel2006} in the Liouville space of the
statistical operators and the complete space spanned by the
operators of measurable quantities. We also note that the
linearity gives the one-to-one correspondence between the
subspaces of the above mentioned spaces. When such $ \mathcal{PT}
$-symmetric-like behavior is restricted to only a subspace we
refer to \emph{nonconventional $\mathcal{PT}$-symmetric system
behavior.}

Second, we admit in our analysis more general non-Hermitian $
\mathcal{PT} $ -symmetric Hamiltonians. We recall here that, in
Ref.~\cite{Thapliyal2025}, the non-Hermiticity of the investigated
Hamiltonians originated only in the presence of damping and
amplification terms, whose non-Hermiticity was `remedied' by the
presence of the Langevin fluctuating operator forces
\cite{Perina1991,Vogel2006}. This guarantees the system evolution
preserving the bosonic canonical commutation relations. Here, we
consider also the Hamiltonians that describe bosonic systems with
unidirectional coupling between the modes. The reason is that
unidirectional coupling allows to concatenate two bosonic
subsystems such that they keep their original eigenvalues.
Moreover the original subspaces belonging to the same eigenvalues
merge together which results in the increased EDs. This property
gives rise to the method suggested and elaborated in
Refs.~\cite{Zhong2020,Wiersig2022} that provides QEPs with
higher-order EDs. {Realization of unidirectional propagation
is based upon ring resonators ~\cite{Zhong2020,Wiersig2022}.}
However, we note that, apart from
a complex experimental realization, unidirectional coupling is
highly non-Hermitian and violates reciprocity of physical
processes. Nevertheless, up to our best knowledge, this is the
only method for reaching QEPs and QHPs with high-order EDs for
open bosonic systems under general conditions. {However, as we
show here, the quantum-mechanical consistency of the models with
unidirectional coupling is guaranteed only for short times.}
Higher-order QEPs using unidirectional coupling were already
realized in \cite{Wang2019b}.

{We note that if quantum field-amplitude fluctuations are
completely neglected and pure states are considered, higher-order
EPs in the Hamiltonian spectra can be observed relatively easily.}
For example, higher-order EPs were predicted in optomechanical
\cite{Jing2017} and cavity magnonic systems \cite{Zhang2019a},
those described by the Bose-Hubbard model \cite{Graefe2008}, or
photonic structures \cite{Teimourpour2014,Znojil2018}.
Higher-order EDs were also studied in
Refs.~\cite{Mandal2021,Delplace2021}. They play significant role
in amplification \cite{Zhong2018} and sensing \cite{Hodaei2017} as
well as speeding up entanglement generation \cite{Li2023}.
{However, the dynamics of open quantum systems brings new features
and qualitatively modifies the systems' behavior
\cite{Minganti2019,Minganti2020}.}

Increasing complexity of the bosonic systems also poses the
question about identification of inherited QEPs and QHPs and the
determination of their degeneracies. Whereas simple bosonic
systems allow for analytical derivation of the eigenvalues and
eigenvectors of their dynamical matrices, more complex bosonic
systems admit only numerical treatment. In this case, we may
numerically decompose a given dynamical matrix into its Jordan
form that directly reveals QEPs and QHPs with their degeneracies.
Alternatively, we may add a little $ \epsilon $ perturbation to
any element of the dynamical matrix that can remove both EDs and
DDs and even allow for distinguishing ED and DD.

Some of the effects related to the presence of QEPs (quantum
exceptional points) and QHPs (quantum hybrid points) with
higher-order EDs (exceptional degeneracies) and DDs (diabolical
degeneracies) are observed also in the behavior of higher-order
FOMs (field-operator moments)
\cite{Perina1991,PerinaJr1995,PerinaJr2022a}. We note that we
refer to genuine QEPs and QHPs in the case of higher-order FOMs.
This originates from the fact that the dynamics of $ n $th-order
FOMs is built, in certain sense, as a `multiplied dynamics' of
first-order FOMs with their inherited QEPs and QHPs. This
`multiplied dynamics' then naturally contains the genuine QEPs and
QHPs with higher-order EDs and DDs, as it was discussed in
\cite{PerinaJr2022a}. In general, genuine QEPs with ED orders up
to $ n $th power of ED orders of inherited QEPs of the first-order
FOMs are expected in the dynamics of $ n $th-order FOMs {(for
examples, see Ref.~\cite{PerinaJr2022a}).} The structure of
genuine QEPs and QHPs in the dynamics of higher-order FOMs
intimately depends on that of the inherited QEPs and QHPs found
for the first-order FOMs. For this reason, we have explicitly
revealed the structure of genuine QEPs and QHPs belonging to the
second-order FOMs for the bosonic systems in the tables of
Appendix~B of Ref.~\cite{Thapliyal2025}. They explicitly elucidate
the relation between EDs and DDs of these genuine QEPs and QHPs
and degeneracies of inherited QEPs and QHPs. We note that the
induced QEPs and QHPs, which were introduced in
Ref.~\cite{PerinaJr2022a}, have their origin in the existence of
identical or similar (related by the field commutation relations)
FOMs in the formal construction of higher-order FOMs spaces and
they further increase the multiplicity of spectral degeneracies.
Nevertheless, they do not enrich the system dynamics. There exist
some general properties of EDs and DDs of genuine QEPs and QHPs of
bosonic systems independent of their configuration that we address
here to complete the analysis of specific bosonic systems.

The paper is organized as follows. Section~\ref{sec2} is devoted
to bosonic systems exhibiting nonconventional
$\mathcal{PT}$-symmetric dynamics, as an example, four-mode
systems are analyzed. Section~\ref{sec3} contains the analysis of
bosonic systems with unidirectional coupling in linear
configurations involving in turn from two to six modes. The
dynamics of the two-mode bosonic system with unidirectional
coupling and its applicability are analyzed in Sec.~\ref{sec4}. A
general analysis of genuine and induced QHPs in the dynamics of
arbitrary-order FOMs is given in Sec.~\ref{sec5}.
Section~\ref{sec6} brings conclusions. {Tables summarizing
QEPs and QHPs---along with their eigenvalue EDs and DDs---for
first- and second-order FOM dynamics are given in
Appendix~\ref{AppA}. Appendix~\ref{AppB} presents the numerical
methods used to identify these singularities and their
degeneracies. In Appendix~\ref{AppC}, we analyze the properties of
the Langevin operator forces in the unidirectional-coupling model,
while Appendix~\ref{AppD} details the statistical characteristics
of the two-mode bosonic system under unidirectional coupling.}

\section{Bosonic systems with bidirectional coupling and nonconventional $\mathcal{PT}$-symmetric dynamics}
\label{sec2}

\begin{figure}  
 \begin{centering}
  \includegraphics[width=0.95\hsize]{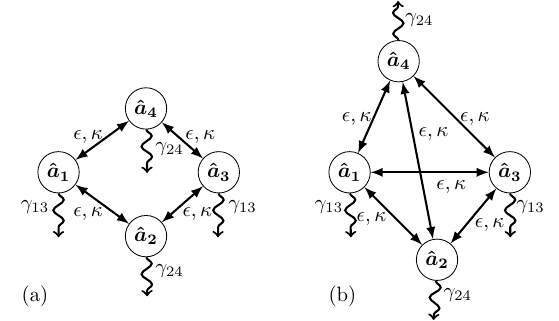}
 \end{centering}

 \caption{Schematic diagrams of the four-mode bosonic systems in (a) circular and (b) tetrahedral configurations
 {that exhibit quantum exceptional points (QEPs) and quantum hybrid points (QHPs) with
 various exceptional degeneracies (EDs) and diabolical
 degeneracies (DDs) observed in their nonconventional $\mathcal{PT}$-symmetric dynamics of field-operator
 moments (FOMs) of different orders.}
 Strengths $ \epsilon $ and $ \kappa $
 characterize, respectively, the linear and nonlinear coupling
 between the modes, while $ \gamma $, with subscripts indicating the mode number(s), are
 the damping or amplification rates, and
 the annihilation operators $ \hat{a} $ identify the mode number via their subscripts.}
\label{fig1}
\end{figure}

When seeking for QEPs and QHPs in simple bosonic systems, we have
observed the situations in which the behavior typical for
$\mathcal{PT}$-symmetric systems occurs only in certain subspaces
of the whole space spanned by the field operators and their
moments. We speak about nonconventional $\mathcal{PT}$-symmetric
dynamics in these subspaces. We note that we may alternatively
specify the corresponding subspaces in the Liouville space of
statistical operators \cite{PerinaJr1995}. As the conditions for
the observation of nonconventional $\mathcal{PT}$-symmetric
dynamics are less restrictive than those required for the usual
$\mathcal{PT}$-symmetric dynamics, we analyze here simple bosonic
systems exhibiting this form of dynamics from the point of view of
the occurrence of higher-order QEPs and QHPs. In the following, we
consider in turn four-mode bosonic systems in their circular and
tetrahedral configurations (see Fig.~\ref{fig1}).

\subsection{Circular configuration}

The Hamiltonian $ \hat{H}_{4,{\rm c}} $ of a four-mode bosonic
system in the circular configuration depicted in
Fig.~\ref{fig1}(a) takes the following form :
\begin{eqnarray}  
 \hat{H}_{4,{\rm c}} &=&  \left[ \hbar \epsilon \hat{a}_{1}^{\dagger}\hat{a}_{2}
   + \hbar \epsilon \hat{a}_{2}^{\dagger}\hat{a}_{3}
   + \hbar \epsilon \hat{a}_{3}^{\dagger}\hat{a}_{4}
   + \hbar \epsilon \hat{a}_{4}^{\dagger}\hat{a}_{1}  + \hbar \kappa \hat{a}_{1}\hat{a}_{2} \right. 
   \nonumber \\
  & &  \left. 
  + \hbar \kappa \hat{a}_{2}\hat{a}_{3} + \hbar \kappa
   \hat{a}_{3}\hat{a}_{4}  +\hbar \kappa \hat{a}_{4}\hat{a}_{1} \right] +
   \textrm{H.c.}
\label{1}   
\end{eqnarray}
where  $\hat{a}_{j} $ ($ \hat{a}_{j}^{\dagger} $) for $
j=1,\ldots, 4 $ denotes the annihilation (creation) operator of
the $j$th mode, $ \epsilon$ ($\kappa$) is the linear (nonlinear)
coupling strength between the modes \cite{Boyd2003}. Symbol
H.c.~replaces the Hermitian-conjugated terms. Damping or
amplification of mode $ j $ is described by the damping
(amplification) rate $ \gamma_j $ and the corresponding Langevin
stochastic operator forces, $ \hat{L}_j $ and $ \hat{L}_j^\dagger
$ that occur in the dynamical Heisenberg-Langevin equations
written below in Eq.~(\ref{2}). The Langevin stochastic operator
forces are assumed to have the Markovian and Gaussian properties
specific to the damping and amplification processes
\cite{PerinaJr2022a}. {In Eq.~(\ref{49}), we present an
example of first- and second-order correlation functions of the
Langevin forces associated with damping (in mode 1) and
amplification (in mode 2).} Their presence in the
Heisenberg-Langevin equations guarantees the fulfillment of the
field-operator commutation relations. Moreover the properties of
the Langevin stochastic operator forces are related to the damping
(amplification) rates via the fluctuation-dissipation theorems
\cite{Vogel2006,Perina1991}.

The Heisenberg-Langevin equations corresponding to the Hamiltonian
$ \hat{H}_{4,{\rm c}} $ in Eq.~(\ref{1}) are written in the form:
\begin{eqnarray}   
 \frac{d\hat{\bm{a}}}{dt} & = & -i \bm{M^{(4)}_{\rm c}} \hat{\bm{a}} +\hat{\bm{L}},
\label{2}
\end{eqnarray}
where the vectors $ \hat{\bm{a}} $ of field operators and $
\hat{\bm{L}} $ of the Langevin operator forces are given as
$\hat{\bm{a}}=\left[\hat{\bm{a}}_{\bm{1}}, \hat{\bm{a}}_{\bm{2}},
\hat{\bm{a}}_{\bm{3}}, \hat{\bm{a}}_{\bm{4}}  \right]^T
\equiv\left[\hat{a}_{1},
\hat{a}_{1}^{\dagger},\hat{a}_{2},\hat{a}_{2}^{\dagger},\hat{a}_{3},\hat{a}_{3}^{\dagger},\hat{a}_{4},\hat{a}_{4}^{\dagger}
\right]^T$ and $\hat{\bm{L}}=\left[\hat{L}_{1},
\hat{L}_{1}^{\dagger},\hat{L}_{2},\hat{L}_{2}^{\dagger},\hat{L}_{3},\hat{L}_{3}^{\dagger},\hat{L}_{4},\hat{L}_{4}^{\dagger}
\right]^T$. The dynamical {$ 8\times 8 $} matrix $
\bm{M^{(4)}_{\rm c}} $ introduced in Eq.~(\ref{2}) is derived in
the form
\begin{eqnarray}  
 \bm{M^{(4)}_{\rm c}}  & = & \left[
  \begin{array}{cccc}
   -i \bm{\tilde{\gamma}_1} & \bm{\xi} & \bm{0} &  \bm{\xi}\\
   \bm{\xi} & -i \bm{\tilde{\gamma}_2} & \bm{\xi}& \bm{0}\\
   \bm{0}  & \bm{\xi} & -i \bm{\tilde{\gamma}_3} &\bm{\xi}\\
    \bm{\xi}  & \bm{0}  &\bm{\xi} & -i \bm{\tilde{\gamma}_4}\\
  \end{array}
 \right].
\label{3}   
\end{eqnarray}
The $ 2\times 2 $ submatrices $ \bm{\tilde{ \gamma}_j} $, $
j=1,\dots,4 $, and $ \bm{\xi} $ occurring in Eq.~(\ref{3}) are
defined as:
\begin{eqnarray}   
 \bm{\tilde{\gamma}_j}=\left[
  \begin{array}{cc}
   \gamma_j/2 & 0 \\
    0 & \gamma_j/2  \\
  \end{array} \right], \hspace{5mm}
 \bm{\xi}=\left[
  \begin{array}{cc}
   \epsilon & \kappa \\
   -\kappa & -\epsilon \\
  \end{array} \right],
  \label{4}    
\end{eqnarray}
and $\gamma_j$ stands for the damping or amplification rate of the
mode $j$.

The submatrices $ \bm{\tilde{ \gamma}_j} $, $ j=1,\dots,4 $, and $
\bm{\xi} $ in Eq.~(\ref{4}) can simultaneously be diagonalized. As
the submatrices $ \bm{\tilde{ \gamma}_j} $ are linearly
proportional to the identity matrix, the appropriate
transformation does not change their form. On the other hand,
it yields two eigenvalues $ \lambda^{\xi}_{1,2} $ of the matrix $
\bm{\xi} $ which we identify by a new variable $ \xi $. In this
diagonalized form of the above submatrices, the original $ 8\times
8 $ dynamical matrix $ \bm{M^{(4)}_{\rm c}} $ attains the form
\begin{eqnarray}  
 \bm{M^{(4)}_{\rm c} }
=  \left[
  \begin{array}{cccccccc}
   \Blue{-i {\gamma}_1/2} & 0    & \Blue{\lambda^{\xi}_{1}} & 0  & \Blue{0} & 0 & \Blue{ \lambda^{\xi}_{1}} & 0 \\
   0 & \Green{ -i {\gamma}_1/2} &  0 & \Green{\lambda^{\xi}_{2} }  & 0 & \Green{0} &  0 & \Green{\lambda^{\xi}_2} \\
   \Blue{\lambda^{\xi}_{1}} & 0 & \Blue{-i {\gamma}_2/2} & 0    & \Blue{\lambda^{\xi}_{1}} & 0 &\Blue{ 0} &  0 \\
   0 & \Green{ \lambda^{\xi}_{2}} &  0 & \Green{ -i {\gamma}_2/2} &  0 & \Green{ \lambda^{\xi}_{2}}   & 0 &\Green{  0} \\
   \Blue{0} &  0 & \Blue{\lambda^{\xi}_{1}} & 0 & \Blue{ -i {\gamma}_3/2} & 0 & \Blue{\lambda^{\xi}_{1}} & 0 \\
   0 & \Green{ 0} & 0 & \Green{\lambda^{\xi}_{2}} & 0 & \Green{-i {\gamma}_3/2} & 0 & \Green{\lambda^{\xi}_{2}} \\
   \Blue{\lambda^{\xi}_{1}} & 0  & \Blue{0} & 0 &   \Blue{ \lambda^{\xi}_{1}} & 0 & \Blue{ -i {\gamma}_4/2} & 0\\
   0 & \Green{\lambda^{\xi}_{2}}    & 0 & \Green{0} &    0 & \Green{\lambda^{\xi}_{2}}  & 0 & \Green{-i
   {\gamma}_4/2}
  \end{array}
 \right] .
\nonumber   
\end{eqnarray}
{In this form, the matrix $ \bm{M^{(4)}_{\rm c}} $ is
decomposed into the direct sum of two $ 4\times 4 $ matrices
belonging to \Blue{$ \xi = \lambda^{\xi}_{1} $} and \Green{$ \xi =
\lambda^{\xi}_{2} $}.} This allows us to effectively consider the
dynamical matrix $ \bm{M^{(4)}_{\rm c}} $ in its $ 4\times 4 $
block structure as if there are just numbers instead of $ 2\times
2 $ submatrices. This effectively halves the dimension of the
diagonalization procedure, allowing us to derive important
analytical results.

To reveal QEPs and QHPs, inspired by $ \mathcal{PT} $-symmetry and
the results of Part I of Ref.~\cite{Thapliyal2025}, we apply the
conditions
\begin{equation}  
  \gamma_1 =\gamma_3 \equiv \gamma_{13}, \hspace{5mm} \gamma_2 =\gamma_4 \equiv
  \gamma_{24},
\label{5}
\end{equation}
in the dynamical {$ 4\times 4 $} matrix $ \bm{ M^{(4)}_{\rm
c}} $ in Eq.~(\ref{3}). Then, we reveal its eigenvalues $
\lambda_{j}^{M^{(4)}_{\rm c}} $:
\begin{eqnarray}   
 \lambda_{1}^{M^{(4)}_{\rm c}} &=& -i \gamma_{13}, \nonumber \\
 \lambda_{2}^{M^{(4)}_{\rm c}} &=& -i \gamma_{24},\nonumber \\
 \lambda_{3,4}^{M^{(4)}_{\rm c}} &=& -i \gamma_{+} \mp \beta.
\label{6}   
\end{eqnarray}
The corresponding eigenvectors are derived as follows:
\begin{eqnarray}   
 \bm{y_{1}^{M^{(4)}_{\rm c}} } &=& \left[-1, 0,1, 0 \right]^T , \nonumber \\
 \bm{y_{2}^{M^{(4)}_{\rm c}} } &=& \left[0,-1, 0,1 \right]^T, \nonumber \\
 \bm{y_{3}^{M^{(4)}_{\rm c}} } &=& \left[-\frac{ 2\xi}{\chi^{*}}, 1,-\frac{ 2\xi}{\chi^{*}} , 1 \right]^T,
   \nonumber \\
 \bm{y_{4}^{M^{(4)}_{\rm c}} } &=& \left[\frac{ 2\xi}{\chi}, 1,\frac{ 2\xi}{\chi} , 1
 \right]^T,
\label{7}   
\end{eqnarray}
where $ \chi= i\gamma_{+} + \beta $, $ \beta^2=4\xi^2-
\gamma_{-}^2$, and $ 4 \gamma_{\pm} = \gamma_{13} \pm \gamma_{24}
$.

\begin{figure}  
 \centerline{{\small (a)}  \includegraphics[width=0.4\hsize]{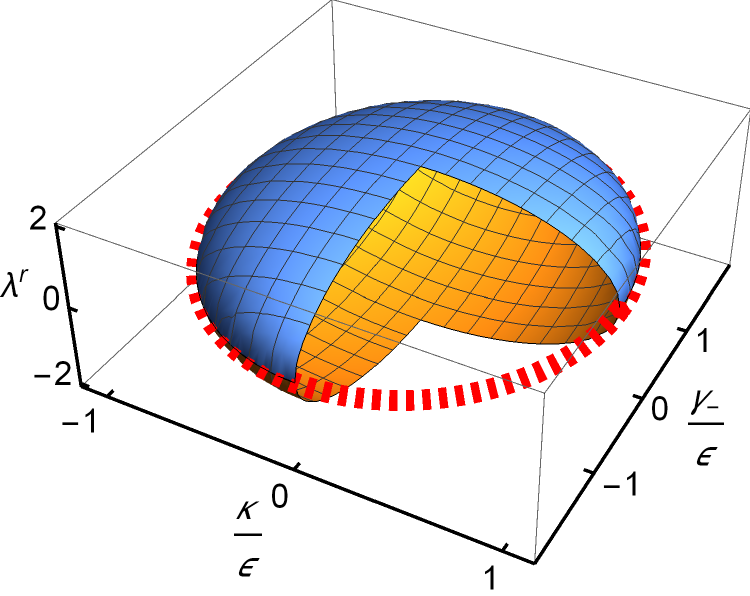}
 {\small (b)}  \includegraphics[width=0.4\hsize]{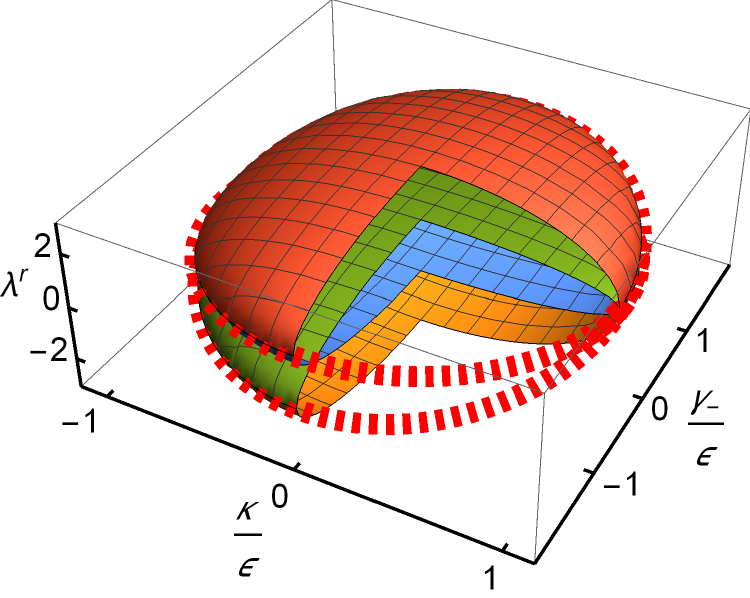} }

 \caption{Real parts $ \lambda^{\rm r} $ of the eigenvalues (a) $ \lambda_{3,4}^{ M^{(4)}_{\rm c}} $
  of the matrix $ \bm{M^{(4)}_{\rm c}} $, given in Eq.~(\ref{6}), for the
  four-mode bosonic system in the circular configuration
  with different damping and/or amplification rates of neighbor
  modes and
  (b) $ \lambda_{3,4}^{ M^{(4)}_{\rm t}} $ of the matrix $ \bm{M^{(4)}_{\rm t}} $,
  given in Eq.~(\ref{17}), for $ \xi=\pm\zeta $ for the four-mode bosonic
  system in the tetrahedral configuration with the same damping and/or amplification rates
  of neighbor modes are drawn in the parameter space
  $ (\kappa/\epsilon,\gamma_-/\epsilon) $. {The dashed red curves indicate the values at the positions of the
  QHPs of the four-mode bosonic systems as given by Eq.~(\ref{8}).} }
\label{fig2}
\end{figure}

If $\beta=0$ then $ \lambda_{3}^{M^{(4)}_{\rm c}} =
\lambda_{4}^{M^{(4)}_{\rm c}} $ and also $ \bm{y_{3}^{M^{(4)}_{\rm
c}} } = \bm{y_{4}^{M^{(4)}_{\rm c}} } $. We note that the
eigenvalues $ \lambda_{3}^{M^{(4)}_{\rm c}} $ and $
\lambda_{4}^{M^{(4)}_{\rm c}} $ share also their imaginary parts,
which is important for the observation of $ \mathcal{PT}
$-symmetric-like dynamics in a suitable interaction frame
\cite{Chimczak2023}. Provided that the system initial conditions
are chosen such that only the eigenvalues $
\lambda_{3}^{M^{(4)}_{\rm c}} $ and $ \lambda_{4}^{M^{(4)}_{\rm
c}} $ determine its dynamics, we observe a second-order QEP. This
QEP changes into a QHP with second-order ED and DD when the $
8\times 8 $ matrix $ \bm{M^{(4)}_{\rm c}} $ is analyzed (see
below). The condition $\beta=0$ assuming $ \xi^2 = \zeta^2
=\epsilon^2-\kappa^2$ [see below Eq.~(\ref{10})] transforms into
the formula
\begin{equation}   
 \frac{\kappa^2}{\epsilon^2}+ \frac{\gamma_{-}^2}{4\epsilon^2}=1
\label{8}
\end{equation}
for an ellipse in the parameter space $
(\kappa/\epsilon,\gamma_-/\epsilon) $ that identifies the
positions of QHPs. Real parts of two eigenvalues $
\lambda_j^{M^{(4)}_{\rm c}} $, $ j=3,4 $, that form QHPs are
plotted in this space in Fig.~\ref{fig2}(a).

The diagonalized $ 8\times 8 $ dynamical matrix $ \bm{
M^{(4)}_{\rm c}} $ is obtained with the help of the eigenvalues
and eigenvectors given in Eqs.~(\ref{6}) and (\ref{7}),
respectively, and the eigenvalues $ \lambda^{\xi}_{1,2} $ and the
eigenvectors $ \bm{y^{\xi}_{1,2}} $ of the matrix $\bm{\xi}$:
\begin{eqnarray} 
 \lambda^{\xi}_{1,2}& =&\mp \zeta,
\label{9}  
\end{eqnarray}
\begin{eqnarray}  
 \bm{y^{\xi}_{1,2}}& =&\left[-\frac{\epsilon \mp \zeta}{\kappa} , 1 \right]^T,
\label{10} 
\end{eqnarray}
where $\zeta=\sqrt{\epsilon^2-\kappa^2}$. {In detail, the
eigenvalues $ \Lambda_{j}^{M^{(4)}_{\rm c}} $ for $ j=1,\ldots,8 $
and the corresponding eigenvectors $ \bm{Y_{j}^{M^{(4)}_{\rm c}}}
$ of the full $ 8\times 8 $ matrix $ \bm{ M^{(4)}_{\rm c}} $ are
obtained using the general scheme applicable to the fields
composed of, in general, $ n $ modes. Relying on the results
applicable to a general $ 2n\times 2n $ dimensional matrix $
\bm{M^{(n)}} $ presented in Part I of Ref.~\cite{Thapliyal2025}
(Appendix~A), for which we have
\begin{eqnarray} 
 \Lambda_{2j-1}^{M^{(n)}} &=&
   {\lambda_{j}^{M^{(n)}}}({\xi}=\lambda^{\xi}_{1}), \nonumber \\
 \Lambda_{2j}^{M^{(n)}} &=& {\lambda_{j}^{M^{(n)}} }
   ({\xi}=\lambda^{\xi}_{2}) ,
\label{11}    \\
 \bm{Y_{2j-1}^{M^{(n)}}} &=& \left[
  \begin{array}{c}
   {y_{j,1}^{M^{(n)}} }({\xi}=\lambda^{\xi}_{1}) \bm{y^{\xi}_{1}} \\
   {y_{j,2}^{M^{(n)}} }({\xi}=\lambda^{\xi}_{1}) \bm{y^{\xi}_{1}} \\
   \ldots \\
   {y_{j,n}^{M^{(n)}} }({\xi}=\lambda^{\xi}_{1}) \bm{y^{\xi}_{1}}
   \end{array} \right], \nonumber \\
 \bm{Y_{2j}^{M^{(n)}}} &=& \left[
  \begin{array}{c}
   {y_{j,1}^{M^{(n)}} }({\xi}=\lambda^{\xi}_{2}) \bm{y^{\xi}_{2}} \\
   {y_{j,2}^{M^{(n)}} }({\xi}=\lambda^{\xi}_{2}) \bm{y^{\xi}_{2}} \\
   \ldots \\
   {y_{j,n}^{M^{(n)}} }({\xi}=\lambda^{\xi}_{2}) \bm{y^{\xi}_{2}} \end{array}\right] ,
\label{12}   \\
   & & \hspace{2mm} j=1,\ldots,n, \nonumber
\end{eqnarray}
the appropriate results are obtained assuming $ n= 4 $ and $
\bm{M^{(n)}} = \bm{M^{(4)}_{\rm c}} $.}

{In the basis in which the $ 8\times 8 $ dynamical matrix $
\bm{ M^{(4)}_{\rm c}} $ attains its diagonal form, the system
dynamics is described by the new field operators
$\hat{\bm{b}}=\left[\hat{b}_{1},\hat{b}_{1}^{\dagger},\hat{b}_{2},\hat{b}_{2}^{\dagger},\hat{b}_{3},\hat{b}_{4},\hat{b}_{3}^{\dagger},
\hat{b}_{4}^{\dagger} \right]^T$ arising in this diagonalization.
We note that the diagonalization and introduction of the new field
operators in the two-mode bosonic system is explicitly given in
Eq.~(14) of Sec.~II of Part~I of Ref.~\cite{Thapliyal2025}. We
also note that the order of elements in the operator vector $
\hat{\bm{b}} $ is given by the numbering of the eigenvalues and
the corresponding diagonalization transform.} If the initial
conditions allow to describe the complete system dynamics in terms
of the field operators $\hat{b}_{3}$, $\hat{b}_{4}$,
$\hat{b}_{3}^\dagger $, and $\hat{b}_{4}^\dagger $ with equal
damping or amplification rate $ \gamma_{+}$ the first- and
second-order FOMs exhibit in their dynamics QEPs and QHPs
summarized in Tab.~\ref{tab2} in Appendix~\ref{AppA}.

\subsection{Tetrahedral configuration}

In the tetrahedral configuration depicted in Fig.~\ref{fig1}(b),
the Hamiltonian $ \hat{H}_{4,{\rm t}} $ of four-mode bosonic
system attains the form:
\begin{eqnarray}  
 \hat{H}_{4,{\rm t}} &=& \left[ \hbar \epsilon \hat{a}_{1}^{\dagger}\hat{a}_{2}
 + \hbar \epsilon \hat{a}_{1}^{\dagger}\hat{a}_{3}
   + \hbar \epsilon \hat{a}_{1}^{\dagger}\hat{a}_{4}
   + \hbar \epsilon \hat{a}_{2}^{\dagger}\hat{a}_{3}
   + \hbar \epsilon \hat{a}_{2}^{\dagger}\hat{a}_{4} \right. \nonumber \\
  & &
   + \hbar \epsilon \hat{a}_{3}^{\dagger}\hat{a}_{4} + \hbar \kappa \hat{a}_{1}\hat{a}_{2} + \hbar \kappa \hat{a}_{1}\hat{a}_{3}
   + \hbar \kappa \hat{a}_{1}\hat{a}_{4} + \hbar \kappa \hat{a}_{2}\hat{a}_{3}   \nonumber \\
  & & \left. + \hbar \kappa \hat{a}_{2}\hat{a}_{4} + \hbar \kappa
   \hat{a}_{3}\hat{a}_{4}  \right]  +
   \textrm{H.c.}
\label{13}   
\end{eqnarray}

The Heisenberg-Langevin equations corresponding to the Hamiltonian
$ \hat{H}_{4,{\rm t}} $ are derived in the form:
\begin{eqnarray}  
 \frac{d\hat{\bm{a}}}{dt} & = & -i \bm{M^{(4)}_{\rm t}} \hat{\bm{a}} +\hat{\bm{L}}
\label{14}
\end{eqnarray}
using the following dynamical matrix $ \bm{M^{(4)}_{\rm t}} $:
\begin{eqnarray}  
 \bm{M^{(4)}_{\rm t}}  & = & \left[
  \begin{array}{cccc}
   -i \bm{\tilde{\gamma}_1} & \bm{\xi} & \bm{\xi} & \bm{\xi}\\
   \bm{\xi} & -i \bm{\tilde{\gamma}_2} & \bm{\xi}& \bm{\xi}\\
   \bm{\xi}  & \bm{\xi} & -i \bm{\tilde{\gamma}_3} &\bm{\xi}\\
   \bm{\xi}  & \bm{\xi}  &\bm{\xi} & -i \bm{\tilde{\gamma}_4}\\
  \end{array}
 \right].
\label{15}   
\end{eqnarray}

In seeking QEPs, we assume equal damping and/or amplification
rates of modes 1 and 2, and also of modes 3 and 4:
\begin{equation}   
 \gamma_1 =\gamma_2 \equiv \gamma_{12}, \hspace{5mm}
 \gamma_3 =\gamma_4 \equiv \gamma_{34}.
\label{16}
\end{equation}
We note that, due to the symmetry, identical results are obtained
when assuming equal damping and/or amplification rates of modes 1
and 3 and also of modes 2 and 4 [compare Eq.~(\ref{5})].

Under these conditions, diagonalization of the {$ 4\times 4 $}
dynamical matrix $\bm{M^{(4)}_{\rm t} }$ in Eq.~(\ref{15}) leaves
us with the following eigenvalues:
\begin{eqnarray}  
  \lambda_{1}^{M^{(4)}_{\rm t}} &=& -i \gamma_{12}- \xi, \nonumber \\
 \lambda_{2}^{M^{(4)}_{\rm t}} &=& -i \gamma_{34} -\xi,\nonumber \\
 \lambda_{3,4}^{M^{(4)}_{\rm t}} &=& - i \gamma_{+} + \xi \mp
 \beta.
\label{17}   
\end{eqnarray}
The corresponding eigenvectors are written as:
\begin{eqnarray}   
 \bm{y_{1}^{M^{(4)}_{\rm t}} } &=& \left[-1, 1,0, 0 \right]^T , \nonumber \\
 \bm{y_{2}^{M^{(4)}_{\rm t}} } &=& \left[0,0, -1,1 \right]^T, \nonumber \\
 \bm{y_{3}^{M^{(4)}_{\rm t}} } &=& \left[1-\frac{2i \gamma_{-}}{\chi_{-}}, 1-\frac{2i \gamma_{-}}{\chi_{-}} ,1, 1 \right]^T,
   \nonumber \\
 \bm{y_{4}^{M^{(4)}_{\rm t}} } &=& \left[1-\frac{2i \gamma_{-}}{\chi_{+}}, 1-\frac{2i \gamma_{-}}{\chi_{+}} ,1, 1
 \right]^T,
\label{18}
\end{eqnarray}
and $\chi_{\pm}=i \gamma_{-}\pm\beta +2\xi $, $ \beta^2=4\xi^2-
\gamma_{-}^2$, and $ 4 \gamma_{\pm} = \gamma_{12} \pm \gamma_{34}
$.

Provided that  $\beta=0$, we have $\lambda_{3}^{M^{(4)}_{\rm t}} =
\lambda_{4}^{M^{(4)}_{\rm t}} $ and $\bm{y_{3}^{M^{(4)}_{\rm t}} }
= \bm{y_{4}^{M^{(4)}_{\rm t}} } $ as $\chi_{-}=\chi_{+}$. As the
imaginary parts of eigenvalues $  \lambda_{1,2}^{M^{(4)}_{\rm t}}
$ differ from those of $ \lambda_{3,4}^{M^{(4)}_{\rm t}} $, the
system can exhibit only the non-conventional
$\mathcal{PT}$-symmetric dynamics: If the system initial
conditions are such that only the eigenvalues $
\lambda_{3}^{M^{(4)}_{\rm t}} $ and $ \lambda_{4}^{M^{(4)}_{\rm
t}} $ suffice in describing its dynamics, we observe a
second-order QEP for the $ 4\times 4 $ dynamical matrix
$\bm{M^{(4)}_{\rm t} }$. As the eigenvalues $
\lambda_j^{M^{(4)}_{\rm t}} $ in Eq.~(\ref{17}) show the linear
dependence on $ \xi $, the diabolical second-order degeneracy of
the $ 8\times 8 $ dynamical matrix $\bm{M^{(4)}_{\rm c} }$,
originating in the form of the eigenvalues $
\lambda_j^{M^{(4)}_{\rm c}} $ in Eq.~(\ref{6}), is not observed in
the tetrahedral configuration. Instead, for $\beta=0$, we find one
second-order QEP for $ \xi = \zeta $  and another second-order QEP
for $ \xi = - \zeta $. These QEPs occur at the positions described
in Eq.~(\ref{8}) in the parameter space $
(\kappa/\epsilon,\gamma_-/\epsilon) $. Real parts of four
eigenvalues $ \lambda_{3,4}^{M^{(4)}_{\rm t}} $ for $ \xi=\pm\zeta
$ that build two QEPs are drawn in this space in
Fig.~\ref{fig2}(b).

In the basis with the diagonal $ 8\times 8 $ dynamical matrix $
\bm{ M^{(4)}_{\rm t}} $, the system dynamics is described by the
new field operators
$\hat{\bm{b}}=\left[\hat{b}_{1},\hat{b}_{1}^{\dagger},\hat{b}_{2},\hat{b}_{2}^{\dagger},\hat{b}_{3},\hat{b}_{4},\hat{b}_{4}^{\dagger},
\hat{b}_{3}^{\dagger} \right]^T$. The corresponding eigenvalues
and eigenvectors are discussed in general in Appendix~A of
Ref.~\cite{Thapliyal2025}. The first- and second-order FOMs
exhibit in their dynamics QEPs and QHPs provided in
Tab.~\ref{tab2} of Appendix~\ref{AppA}.

\section{Concatenated bosonic systems with unidirectional coupling:
Higher-order quantum exceptional points on demand} \label{sec3}

The analysis of $\mathcal{PT}$-symmetric bosonic systems with up
to five modes in their linear, circular, tetrahedron, and pyramid
configurations presented above and in Part I of
Ref.~\cite{Thapliyal2025} revealed only the inherited QEPs and
QHPs with second- and third-order EDs. That is why, we extend our
analysis to more general non-Hermitian Hamiltonians that involve
unidirectional coupling between the modes. Whereas the
non-Hermiticity of the above discussed systems is given solely by
the presence of damping and/or amplification, the Hamiltonians
with unidirectional coupling are non-Hermitian per se. Despite
their non-Hermiticity, they have direct {effective} physical
implementations based upon counter-directional field propagation
and mutual scattering \cite{Zhong2020,Wiersig2022} {or
nonlinear Kerr interaction \cite{Wang2019}.} Moreover, it has been
shown in Ref.~\cite{McDonald2020} that exponential improvement of
measurement precision can be reached in QEPs in systems with
unidirectional coupling.

It was shown in Refs.~\cite{Zhong2020,Wiersig2022} that using a
specific unidirectional coupling of two field modes belonging to
different $\mathcal{PT}$-symmetric systems with QEPs, the combined
system exhibits a QEP with ED given as the sum of those of the
constituting systems. This opens the door for observing inherited
QEPs with EDs of orders higher than three.

\begin{figure*}  
 \begin{centering}
  \includegraphics[width=0.85\hsize]{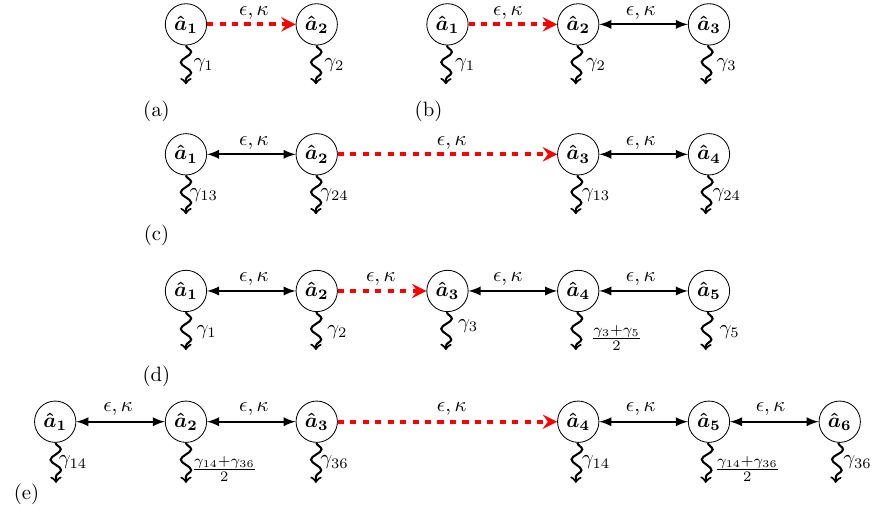}
 \end{centering}

 \caption{Schematic diagrams of bosonic systems composed of two subsystems mutually coupled by unidirectional
  coupling and having: (a) two, (b) three, (c) four, (d) five, and (e) six modes
  in typical linear configurations {that exhibit quantum exceptional points (QEPs) and quantum hybrid points (QHPs) with
  various exceptional degeneracies (EDs) and diabolical degeneracies (DDs) observed in the dynamics of field-operator
  moments (FOMs) of different orders.} The coupling strengths
  $ \epsilon $ and $ \kappa $ characterize both unidirectional {(red dashed
  arrows)}
  and bidirectional {(back full double arrows)} coupling between modes, $ \gamma $, with subscripts indicating the mode
  number(s), give the damping or amplification rates, and
  the annihilation operators $ \hat{a} $ identify the mode number via their subscripts.}
\label{fig3}
\end{figure*}

To demonstrate the method, let us consider two bosonic modes that
are unidirectionally coupled via the matrix $ \bm{\xi} $ defined
in Eq.~(\ref{4}) [see the scheme in Fig.~\ref{fig3}(a)].
As we demonstrate below, this unidirectional coupling creates a
QEP. The corresponding Heisenberg-Langevin equations take the form
\begin{eqnarray}   
 \frac{d\hat{\bm{a}}}{dt} & = & -i \bm{M^{(1+1)}_{\rm u}} \hat{\bm{a}}
 +\hat{\bm{L}},
\label{19}
\end{eqnarray}
where the vectors $ \hat{\bm{a}} $ of field operators and $
\hat{\bm{L}} $ of the Langevin operator forces are given as
$\hat{\bm{a}}=\left[\hat{\bm{a}}_{\bm{1}}, \hat{\bm{a}}_{\bm{2}}
\right]^T \equiv\left[\hat{a}_{1},
\hat{a}_{1}^{\dagger},\hat{a}_{2},\hat{a}_{2}^{\dagger} \right]^T$
and $\hat{\bm{L}}=\left[\hat{L}_{1},
\hat{L}_{1}^{\dagger},\hat{L}_{2},\hat{L}_{2}^{\dagger}\right]^T$.
The properties of the Langevin operator forces are described in
detail below, i.e. in Eq.~(\ref{49}).

The dynamical {$ 4\times 4 $} matrix $ \bm{M^{(1+1)}_{\rm u}}
$ introduced in Eq.~(\ref{19}) and corresponding to unidirectional
coupling of modes is written as
\begin{eqnarray} 
\bm{M^{(1+1)}_{\rm u}}  & = & \left[
 \begin{array}{ccc}
  -i \bm{\tilde{\gamma}_1} & \bm{0}  \\
  \bm{\xi} & -i \bm{\tilde{\gamma}_2} \\
 \end{array} \right],
\label{20}  
\end{eqnarray}
using the damping or amplification $ 2 \times 2 $ submatrices $
\bm{\tilde{\gamma}_j} $, $ j=1,2 $, defined in Eq.~(\ref{4}). The
eigenvalues and eigenvectors of the $ 2\times 2 $ matrix
$\bm{M^{(1+1)}_{\rm u}}$ {with $ 2\times 2 $ submatrices as
its formal elements} are determined as
\begin{equation}  
  \lambda_{1,2}^{M^{(1+1)}_{\rm u}} = -i \gamma_{1,2}/2,
\label{21}   
\end{equation}
and
\begin{eqnarray}   
 \bm{y_{1}^{M^{(1+1)}_{\rm u} } }&=&\left[-\frac{ \gamma_{1}-\gamma_{2}}{2\xi}, 1 \right]^T, \nonumber \\
 \bm{y_{2}^{M^{(1+1)}_{\rm u} } } &=& \left[0 , 1 \right]^T.
\label{22}   
\end{eqnarray}
{To observe QEPs and QHPs, we assume}
\begin{equation}   
 \gamma_{1}=\gamma_{2}.
\label{23}
\end{equation}
Then we have $ \lambda_{1}^{M^{(1+1)}_{\rm u}} =
\lambda_{2}^{M^{(1+1)}_{\rm u}} $ together with $
\bm{y_{1}^{M^{(1+1)}_{\rm u}}} = \bm{y_{2}^{M^{(1+1)}_{\rm u}}} $,
and so we have a QEP with second-order ED. This means that the $
4\times 4 $ matrix $\bm{M^{(1+1)}_{\rm u}}$ obtained after
inserting the $ 2\times 2 $ submatrices $ \bm{\tilde{\gamma}_j} $,
$ j=1,2 $, and $ \bm{\xi} $ into Eq.~(\ref{20}) exhibits a QHP
with second-order ED and DD. It is worth noting that these QHPs
occur independently of the values of the coupling strengths $
\epsilon $ and $ \kappa $, in strike difference to the QHPs found
in the two-mode bosonic system analyzed in Sec.~II of Part I of
Ref.~\cite{Thapliyal2025}.

In the next step we demonstrate the increase of the order of ED of
a QEP by considering a single-mode bosonic system unidirectionally
coupled to a two-mode bosonic system with a QEP [see the scheme in
Fig.~\ref{fig3}(b)]. The {$ 6\times 6 $} dynamical matrix $
\bm{M^{(1+2)}_{\rm u}} $ of the concatenated system is given as:
\begin{eqnarray} 
 \bm{M^{(1+2)}_{\rm u}} & = & \left[
  \begin{array}{ccc}
   -i \bm{\tilde{\gamma}_1} & \bm{0} & \bm{0} \\
   \bm{\xi} & -i \bm{\tilde{\gamma}_2} & \bm{\xi}\\
  \bm{0}  & \bm{\xi} & -i \bm{\tilde{\gamma}_3}\\
 \end{array} \right].
\label{24}    
\end{eqnarray}
{To observe QEPs and QHPs, we assume}
\begin{equation}   
  \gamma_1 = (\gamma_2 + \gamma_3)/2 \equiv 2\gamma_+
\label{25}
\end{equation}
and we arrive at the eigenvalues of the {$ 3\times 3 $ matrix
$ \bm{M^{(1+2)}_{\rm u}} $}:
\begin{eqnarray}   
 \lambda_{1}^{M^{(1+2)}_{\rm u}} &=& -i \gamma_+ , \nonumber \\
 \lambda_{2,3}^{M^{(1+2)}_{\rm u}} &=& -i \gamma_+ \mp \beta,
\label{26}   
\end{eqnarray}
and the corresponding eigenvectors:
\begin{eqnarray}  
 \bm{y_{1}^{M^{(1+2)}_{\rm u}}} &=& \left[\frac{\beta^2}{\xi^2}, \frac{i\gamma_{-}}{\xi}, 1 \right]^T, \nonumber \\
 \bm{y_{2,3}^{M^{(1+2)}_{\rm u}}} &=& \left[0 ,\frac{-i\gamma_{-}\mp \beta }{\xi} , 1
 \right]^T.
\label{27}   
\end{eqnarray}
In Eqs.~(\ref{26}) and (\ref{27}), $ \beta^2 = \xi^2 - \gamma_-^2
$ and $ 4\gamma_{\pm} = \gamma_2 \pm \gamma_3 $. If $ \beta=0 $,
i.e., when the condition
\begin{equation}   
 \frac{\kappa^2}{\epsilon^2}+ \frac{\gamma_{-}^2}{\epsilon^2}=1
\label{28}
\end{equation}
for the constituting two-mode system is fulfilled, we have $
\lambda_{1}^{M^{(1+2)}_{\rm u}} = \lambda_{2}^{M^{(1+2)}_{\rm u}}
= \lambda_{3}^{M^{(1+2)}_{\rm u}} $, and $
\bm{y_{1}^{M^{(1+2)}_{\rm u}}} = \bm{y_{2}^{M^{(1+2)}_{\rm u}}} =
\bm{y_{3}^{M^{(1+2)}_{\rm u}}} $. This identifies a QEP with
third-order ED that emerged from the original QEP with
second-order ED. This means that we have a QHP with third-order ED
and second-order DD in the $ 6\times 6 $ matrix $
\bm{M^{(1+2)}_{\rm u}} $. Their eigenvalues and eigenvectors are
constructed using the general scheme in Eqs.~(\ref{11}) and
(\ref{12}). Introducing the new field operators, in which the
{$ 6\times 6 $} dynamical matrix $ \bm{M^{(1+2)}_{\rm u}} $
attains its diagonal form, we reveal QEPs and QHPs and their
degeneracies appropriate to the dynamics of the first- and
second-order FOMs. They can be found in Tab.~\ref{tab4} of
Appendix~\ref{AppA} derived for a general $ n $-mode bosonic
system exhibiting an inherited QHP with $ n $th-order ED and
second-order DD for $ n>2 $. We note that Tab.~\ref{tab4} of
Appendix~\ref{AppA} applies also to the bosonic systems analyzed
below.

Now, we construct four-, five-, and six-mode bosonic systems by
concatenating the two- and three-mode systems with second- and
third-order QEPs analyzed in Secs.~II and III of
Ref.~\cite{Thapliyal2025}. We begin with combining two two-mode
systems [see the scheme in Fig.~\ref{fig3}(c)] whose dynamical
matrices denoted as $
\bm{M^{(2)}(\tilde{\gamma}_{1},\tilde{\gamma}_{2},\xi)} $ and $
\bm{M^{(2)}(\tilde{\gamma}_{3},\tilde{\gamma}_{4},\xi)} $ are
defined as follows [compare Eq.~(3) of part~I of the paper
\cite{Thapliyal2025}]:
\begin{eqnarray}   
 \bm{M^{(2)}(\tilde{\gamma}_{1},\tilde{\gamma}_{2},\xi)}  & = & \left[
 \begin{array}{cc}
  -i \bm{\tilde{\gamma}_1} & \bm{\xi} \\
  \bm{\xi} & -i \bm{\tilde{\gamma}_2}
 \end{array}
 \right] .
\label{29}  
\end{eqnarray}
The joint {$ 8\times 8 $} dynamical matrix $
\bm{M^{(2+2)}_{\rm u}} $, defined as
\begin{eqnarray}  
 \bm{M^{(2+2)}_{\rm u}}  &=& \left[
  \begin{array}{cc}
   \bm{M^{(2)}(\tilde{\gamma}_{1},\tilde{\gamma}_{2},\xi)} & \bm{0} \\
   \bm{\Upsilon^{(2+2)}} & \bm{M^{(2)}(\tilde{\gamma}_{3},\tilde{\gamma}_{4},\xi)}
  \end{array}\right], \nonumber \\
  & &
\label{30}  
\end{eqnarray}
includes unidirectional coupling between modes 2 and 3 described
by the {$ 4\times 4 $} submatrix $ \bm{\Upsilon^{(2+2)}} $:
\begin{equation}   
 \bm{\Upsilon^{(2+2)}}=\left[\begin{array}{cc}
   \bm{0} & \bm{\xi} \\
   \bm{0} & \bm{0} \end{array}\right];
\label{31}
\end{equation}
{where the symbol $ \bm{0} $ denotes the {$ 2\times 2 $}
null matrix.}

{To reveal QEPs and QHPs, we assume}
\begin{equation}  
 \gamma_1 = \gamma_3 \equiv \gamma_{13}, \hspace{5mm}
 \gamma_2 = \gamma_4 \equiv \gamma_{24},
\label{32}
\end{equation}
and determine the eigenvalues of the $ 4\times 4 $ matrix $
\bm{M^{(2+2)}_{\rm u}} $ as follows:
\begin{eqnarray}   
 \lambda_{1,2}^{M^{(2+2)}_{\rm u}} &=& -i \gamma_{+} \mp \beta,\nonumber\\
 \lambda_{3,4}^{M^{(2+2)}_{\rm u}} &=& -i \gamma_{+} \mp \beta,
\label{33}   
\end{eqnarray}
using $ \beta^2= \xi^2-\gamma_{-}^2$ and $ 4\gamma_{\pm} =
\gamma_{13} \pm \gamma_{24}$. For $ \beta = 0 $, a QEP with
fourth-order ED is found at the positions in the parameter space
given by Eq.~(\ref{28}), i.e., where the constituting two-mode
systems form QEPs with second-order EDs. The fourth-order QEP
implies a QHP with fourth-order ED and second-order DD of the $
8\times 8 $ matrix $ \bm{M^{(2+2)}_{\rm u}} $. It is worth noting
that the additional unidirectional coupling of modes 1 and 4, i.e.
when
$\bm{\bar{\Upsilon}^{(2+2)}}=\left[\begin{array}{cc} \bm{0} & \bm{\xi} \\
\bm{\xi} & \bm{0} \end{array}\right] $, does not change the
eigenvalues in Eq.~(\ref{33}) and also preserves the structure of
eigenvectors with the discussed QEPs and QHPs.

The unidirectional coupling of the two- and three-mode systems
allows to observe a QEP with fifth-order ED [for the
configuration, see Fig.~\ref{fig3}(d)]. The corresponding {$
10\times 10 $} dynamical matrix $ \bm{M^{(2+3)}_{\rm u}} $
combines the two-mode  {$ 4\times 4 $} dynamical matrix $
\bm{M^{(2)}} $ in Eq.~(\ref{29}) and the three-mode  {$ 6\times
6 $} dynamical matrix $ \bm{M^{(3)}} $ [compare Eq.~(21) in
\cite{Thapliyal2025}],
\begin{eqnarray}   
 \bm{M^{(3)}(\tilde{\gamma}_{1},\tilde{\gamma}_{2},\tilde{\gamma}_{3},\xi)} & = & \left[
  \begin{array}{ccc}
   -i \bm{\tilde{\gamma}_1} & \bm{\xi} & \bm{0} \\
   \bm{\xi} & -i \bm{\tilde{\gamma}_2} & \bm{\xi}\\
  \bm{0}  & \bm{\xi} & -i \bm{\tilde{\gamma}_3}\\
  \end{array}
  \right].
\label{34} 
\end{eqnarray}
The  {$ 10\times 10 $} matrix $ \bm{M^{(2+3)}_{\rm u}} $ is
expressed as
\begin{eqnarray}  
 \bm{M^{(2+3)}_{\rm u}}  &=& \left[
  \begin{array}{cc}
   \bm{M^{(2)}(\tilde{\gamma}_{1},\tilde{\gamma}_{2},\xi)} & \bm{0} \\
   \bm{\Upsilon^{(2+3)}} & \bm{M^{(3)}(\tilde{\gamma}_{3},\tilde{\gamma}_{4},\tilde{\gamma}_{5},\xi)}
  \end{array}\right], \nonumber \\
  & &
\label{35}
\end{eqnarray}
assuming the unidirectional coupling between modes 2 and~3:
\begin{equation}   
 \bm{\Upsilon^{(2+3)}}=\left[\begin{array}{cc}
   \bm{0} & \bm{\xi} \\
   \bm{0} & \bm{0} \\
      \bm{0} & \bm{0} \end{array}\right].
\label{36}
\end{equation}
Provided that
\begin{equation}  
 \gamma_1 + \gamma_2 = \gamma_3 + \gamma_5 ,
 \hspace{5mm}  2\gamma_4 = \gamma_3 + \gamma_5,
\label{37}
\end{equation}
we determine the eigenvalues of the $ 5\times 5 $ matrix $
\bm{M^{(2+3)}_{\rm u}} $ as follows:
\begin{eqnarray}   
 \lambda_{1}^{M^{(2+3)}_{\rm u}} &=& -i \gamma_{+}, \nonumber\\
 \lambda_{2,3}^{M^{(2+3)}_{\rm u}} &=& -i \gamma_{+} \mp \bar{\beta},\nonumber\\
 \lambda_{4,5}^{M^{(2+3)}_{\rm u}} &=& -i \gamma_{+} \mp \beta,
\label{38}   %
\end{eqnarray}
where $ \beta^2= \xi^2-\gamma_{-}^2$, $ \bar{\beta}^2=
2\xi^2-\bar{\gamma}_{-}^2$, $ 4\gamma_{\pm} = \gamma_{1} \pm
\gamma_{2} $, and $ 4\bar{\gamma}_{-} = \gamma_{3} - \gamma_{5} $.
Provided that $ \beta = 0 $ and $ \bar{\beta} = 0 $, we find a QEP
with fifth-order ED. These conditions define the positions, which
are specified in Eq.~(\ref{28}) and the following one:
\begin{equation}   
 \frac{\kappa^2}{\epsilon^2}+ \frac{\bar{\gamma}_{-}^2}{2\epsilon^2}=1.
\label{39}
\end{equation}
Both conditions have to be fulfilled simultaneously, which results
in the following condition
\begin{equation}   
 \gamma_- = \bar{\gamma}_-/\sqrt{2} .
\label{40}
\end{equation}
Thus, we reveal the QHPs of the $ 10\times 10 $ matrix $
\bm{M^{(2+3)}_{\rm u}} $ under the conditions, given in
Eqs.~(\ref{37}) and (\ref{40}), observed in the space $
(\kappa/\epsilon, \gamma_-/\epsilon) $ at the positions obeying
Eq.~(\ref{28}).  {We note that the corresponding eigenvalues
and eigenvectors are obtained using the general formulas in
Eqs.~(\ref{11}) and (\ref{12}).}

The last analyzed system is created by unidirectional combining of
two three-mode systems with identical parameters in the
configuration shown in Fig.~\ref{fig3}(e). Its  {$ 12\times 12
$} dynamical matrix, denoted as $ \bm{M^{(3+3)}_{\rm u}} $, is
composed of the two three-mode dynamical matrices $ \bm{M^{(3)}} $
from Eq.~(\ref{34}) and the  {$ 6\times 6 $} coupling matrix $
\bm{\Upsilon^{(3+3)}} $,
\begin{equation}   
 \bm{\Upsilon^{(3+3)}}=\left[\begin{array}{ccc}
   \bm{0} & \bm{0} & \bm{\xi} \\
   \bm{0} & \bm{0} & \bm{0} \\
   \bm{0} & \bm{0} & \bm{0} \end{array}\right].
\label{41}
\end{equation}
It is expressed in the form
\begin{eqnarray}  
 \bm{M^{(3+3)}_{\rm u}}  &=& \left[
  \begin{array}{cc}
   \bm{M^{(3)}(\tilde{\gamma}_{1},\tilde{\gamma}_{2},\tilde{\gamma}_{3},\xi)} & \bm{0} \\
   \bm{\Upsilon^{(3+3)}} & \bm{M^{(3)}(\tilde{\gamma}_{4},\tilde{\gamma}_{5},\tilde{\gamma}_{6},\xi)}
  \end{array}\right] , \nonumber \\
  & &
\label{42}
\end{eqnarray}
in which we assume:
\begin{eqnarray} 
 & \gamma_1 = \gamma_4 \equiv \gamma_{14}; \hspace{5mm} \gamma_3 = \gamma_6 \equiv
 \gamma_{36}, & \nonumber \\
 & 2\gamma_2 = 2\gamma_4 = \gamma_{14} + \gamma_{36}. &
\label{43}
\end{eqnarray}
The eigenvalues of the $ 6\times 6 $ matrix $ \bm{M^{(3+3)}_{\rm
u}} $  {considered in its submatrix form} are determined as:
\begin{eqnarray}   
 \lambda_{1,2}^{M^{(3+3)}_{\rm u}} &=& -i \gamma_{+}, \nonumber\\
 \lambda_{3,4}^{M^{(3+3)}_{\rm u}} &=& -i \gamma_{+} \mp \beta,\nonumber\\
 \lambda_{5,6}^{M^{(3+3)}_{\rm u}} &=& -i \gamma_{+} \mp \beta,
\label{44}   %
\end{eqnarray}
where $ \beta^2= 2\xi^2-\gamma_{-}^2 $ and $ 4\gamma_{\pm} =
\gamma_{14} \pm \gamma_{36} $. It holds $ \beta=0 $ at the
positions in the parameter space $ (\kappa/\epsilon,
\gamma_-/\epsilon) $ that fulfil the condition
\begin{equation}   
 \frac{\kappa^2}{\epsilon^2}+ \frac{\gamma_{-}^2}{2\epsilon^2}=1.
\label{45}
\end{equation}
At these positions we have six identical eigenvalues in
Eq.~(\ref{44}). They indicate a QEP with sixth-order ED that
implies a QHP of sixth-order ED and second-order DD in the
dynamical $ 12\times 12 $ matrix $ \bm{M^{(3+3)}_{\rm u}} $.

In the last three analyzed systems, explicit forms for the
eigenvectors of the (full) dynamical matrices $ \bm{M^{(2+2)}_{\rm
u}} $, $ \bm{M^{(2+3)}_{\rm u}} $, and $ \bm{M^{(3+3)}_{\rm u}} $
were not analyzed because of their complexity. Instead, these
dynamical matrices were diagonalized under the conditions at which
QEPs are expected and the transformed matrices in the Jordan form
with unit elements at the upper diagonal confirmed the presence of
QEPs.  {For details, see Appendix~\ref{AppB}.}

The considered forms of unidirectional coupling matrices $
\bm{\Upsilon^{(2+2)}} $, $ \bm{\Upsilon^{(2+3)}} $, and $
\bm{\Upsilon^{(3+3)}} $ can be replaced by those connecting
different pairs of modes in the constituting systems. This
replacement does not change the eigenvalues as well as the
degeneracies of the eigenvectors that give rise to the observed
QEPs and QHPs. We note that the form of eigenvectors depends on
which pairs of modes are unidirectionally coupled. We can even
include unidirectional coupling of several pairs of modes in the
constituting systems and this property still holds. The only
requirement is that all couplings point out from one subsystem to
the other subsystem.

We note that there exist alternative ways to realize
unidirectional coupling between the constituting bosonic systems.
Contrary to the considered unidirectional coupling that is
described directly via the dynamical matrices, we may consider an
alternative form of unidirectional coupling characterized by the
Hamiltonians
\begin{eqnarray}  
 \hat{H}_{{\rm c-u},1} &=& \hbar \epsilon
  \hat{a}_{i}\hat{a}_{j}^{\dagger} + \hbar \kappa
  \hat{a}_{i}\hat{a}_{j},
\label{46}   \\
 \hat{H}_{{\rm c-u},2} &=& \hbar \epsilon
  \hat{a}_{i}\hat{a}_{j}^{\dagger} + \hbar \kappa
  \hat{a}_{i}^{\dagger}\hat{a}_{j}^{\dagger}.
\label{47}
\end{eqnarray}
They result in the system dynamics similar to that discussed above
and lead to the eigenvalues and eigenvectors with characteristic
EDs and DDs. This improves the feasibility of practical
realizations of such concatenated bosonic systems with
higher-order QEPs and QHPs based on unidirectional coupling.

The method is general, allowing for concatenating simple bosonic
systems into more complex ones that keep QEPs and QHPs of the
original systems using unidirectional coupling of different kinds.
Combing the analyzed bosonic systems with second- and third-order
EDs together, bosonic systems with arbitrary-order EDs can be
achieved.

 {At the end of our analysis of spectral degeneracies in
bosonic systems, based on analytical results, we make the
following remark. For more complex systems, two distinct numerical
approaches can be employed to determine the orders of exceptional
degeneracies. To illustrate the principles and functioning of
these methods, we apply both analytically to the simplest case of
two-mode systems with standard (bidirectional) and unidirectional
couplings, as presented in Appendix~\ref{AppB}.}

\section{Unidirectional coupling and its applicability}
\label{sec4}

In this section, we reveal characteristic features of the bosonic
systems with unidirectional coupling and specify the conditions of
their applicability. We consider the simplest two-mode bosonic
system with unidirectional coupling whose dynamics is described by
the Heisenberg-Langevin equations written in Eq.~(\ref{19}) with
the dynamical matrix $ \bm{M^{(1+1)}_{\rm u}} $ given in
Eq.~(\ref{20}). We assume the most typical configuration of $
\mathcal{PT} $-symmetric systems in which mode 1 is damped and
mode 2 is amplified:
\begin{equation}    
 \gamma_1=-\gamma_2\equiv 2\gamma  .
\label{48}
\end{equation}

The corresponding Langevin fluctuating operator forces embedded in
the vector $ \hat{\bm{L}} $ are modelled by two independent
quantum random Gaussian
processes~\cite{Meystre2007,Agarwal2012,Perinova2019}. This
results in the following correlation functions:
\begin{eqnarray} 
 \langle\hat{L}_1(t)\rangle = \langle\hat{L}_1^\dagger(t)\rangle = 0,  \hspace{3mm}
 \langle\hat{L}_2(t)\rangle = \langle\hat{L}_2^\dagger(t)\rangle = 0,\nonumber \\
 \langle\hat{L}_1^\dagger(t)\hat{L}_1(t')\rangle = 0,
  \hspace{3mm} \langle\hat{L}_1(t)\hat{L}_1^\dagger(t')\rangle =
  2\gamma \delta(t-t'), \nonumber \\
 \langle\hat{L}_2^\dagger(t)\hat{L}_2(t')\rangle = 2\gamma \delta(t-t'),
  \hspace{3mm} \langle\hat{L}_2(t)\hat{L}_2^\dagger(t')\rangle =0.
\label{49}
\end{eqnarray}
The remaining second-order correlation functions are zero. Symbol
$ \delta $ stands for the Dirac function.  {We note that the
applied Markovian statistics of the reservoirs can be extended to
non-Markovian cases by modelling the system's interaction with a
reservoir via an ancilla coupled to a Markovian
bath~\cite{Lin2025}.}

The solution to the stochastic linear differential operator
equations in Eq.~(\ref{19}) is expressed in the form
\cite{PerinaJr2000}
\begin{eqnarray}  
 \hat{\bm{a}}(t) & = &  \bm{P}(t,0) \hat{\bm{a}}(0) +\hat{\bm{F}}(t)
\label{50}
\end{eqnarray}
using the evolution matrix $ \bm{P} $ defined as
\begin{eqnarray}  
 \bm{P}(t,t') &=&  \exp[-i \bm{M^{(1+1)}_{\rm u}}(t-t') ].
 \label{51}
\end{eqnarray}

The fluctuating operator forces $ \hat{\bm{F}} $ introduced in
Eq.~(\ref{50}) are determined along the formula
\begin{eqnarray}  
 \hat{\bm{F}}(t) = \int_0^{t} dt' \bm{P}(t,t')\hat{\bm{L}}(t').
\label{52}
\end{eqnarray}
It implies the following formula for their second-order
correlation functions,
\begin{eqnarray}  
 \langle \hat{\bm{F}}(t) \hat{\bm{F}}^{\dagger \bm{T}}(t) \rangle & = & \int_0^{t} d\tilde{t}\int_0^{t} d\tilde{t}' \nonumber \\
   & & \hspace{-10mm} \bm{P}(t,\tilde{t}) \langle\hat{\bm{L}}(\tilde{t}) \hat{\bm{L}}^{\dagger \bm{T}}(\tilde{t}')\rangle
    \bm{P}^{\dagger \bm{T}}(t,\tilde{t}').
\label{53}
\end{eqnarray}

The solution in Eq.~(\ref{50}) can be recast into a simpler form
written for the annihilation operators $ \hat{a}_1 $ and $
\hat{a}_2 $:
\begin{eqnarray}  
 \left[ \begin{array}{c} \hat{a}_1(t) \\ \hat{a}_2(t) \end{array} \right] & = &
  \bm{U}(t) \left[ \begin{array}{c} \hat{a}_1(0) \\ \hat{a}_2(0) \end{array} \right]
   + \bm{V}(t) \left[ \begin{array}{c} \hat{a}_1^\dagger(0) \\ \hat{a}_2^\dagger(0) \end{array} \right]
   + \left[ \begin{array}{c} \hat{f}_1(t) \\ \hat{f}_2(t) \end{array}
   \right].\nonumber \\
  & &
\label{54}
\end{eqnarray}
The elements of the matrices $ \bm{U} $ and $ \bm{V} $ are defined
as $U_{j,k}(t)=P_{2j-1,2k-1}(t,0) $ and
$V_{j,k}(t)=P_{2j-1,2k}(t,0) $ for $j,k=1,2$, and we also have $
\hat{{f}_j}(t)= \hat{F}_{2j-1}(t) $ for $j=1,2$.

Using the eigenvalues and eigenvectors written in Eqs.~(\ref{9}),
(\ref{10}), (\ref{21}), and (\ref{22}), we arrive at the formulas
specific to our model:
\begin{eqnarray}  
 & \bm{U}(t) =  \left[ \begin{array}{cc}  \mu(t) & 0 \\
   \frac{-i \epsilon s(t)}{\gamma}  & \frac{1}{\mu(t)}
   \end{array}
  \right],  \bm{V}(t) =  \left[
  \begin{array}{cc}
   0 & 0 \\
   \frac{-i \kappa s(t)}{\gamma} & 0
  \end{array}
  \right] , &
 \label{55}  \\
 & \langle \hat{\bm{F}}(t)\hat{\bm{F}}^{\dagger \bm{T}}(t) \rangle =
  \left[ \begin{array}{cc}  \bm{F_1}(t) & \bm{F_{12}}(t)  \\
  \bm{F_{12}^{*T}}(t)  & \bm{F_2}(t) \end{array} \right],&
 \label{56} \\
   & \bm{F_1}(t) = \left[ \begin{array}{cc}  1-\mu^2(t) & 0  \\
   0  & 0 \end{array} \right], \quad \bm{F_{12}} (t) = \frac{i\sigma(t)}{2\gamma}
   \left[ \begin{array}{cc} \epsilon & -\kappa \\
   0  & 0 \end{array} \right], & \nonumber \\
 & \bm{F_2}(t) = \frac{s(t) -2 \gamma t }{2\gamma^2} \left[ \begin{array}{cc}
   \epsilon^2 & -\epsilon\kappa \\
   -\epsilon\kappa  & -\kappa^2 \end{array} \right]
   +  \left(\frac{1}{\mu^2(t)}-1 \right) \left[  \begin{array}{cc}
    0 & 0 \\  0  & 1 \end{array} \right], & \nonumber \\
  &  &
 \label{57}
\end{eqnarray}
where $\mu(t)=\exp(-\gamma t)$, $\sigma(t) = \exp(-2\gamma t) -1+2
\gamma t $, and $ s(t) =\sinh(2\gamma t)$ using the hyperbolic
sinus function.

To check consistency of the model with unidirectional coupling, we
determine the mean values of equal time field-operators
commutation relations. Compared to the usual canonical commutation
relations $\langle [\hat{a}_j(t),\hat{a}_k(t) ] \rangle = 0 $, $
\langle [\hat{a}_j(t),\hat{a}_k^{\dagger}(t) ] \rangle =
\delta_{jk}$, and $ \langle
[\hat{a}_j^\dagger(t),\hat{a}_k^{\dagger}(t) ] \rangle = 0$ for
$j,k=1,2$, the following two relations are found:
\begin{eqnarray}   
 \langle [\hat{a}_2(t),\hat{a}_2^{\dagger}(t) ] \rangle &=& 1+
 \frac{(\epsilon^2 -\kappa^2)\phi(t)}{2\gamma^2},
 \nonumber \\
 \langle [\hat{a}_1(t),\hat{a}_2^{\dagger}(t) ] \rangle &=&
 \frac{-i\epsilon \psi(t)}{2\gamma},
\label{58}
\end{eqnarray}
where $\psi(t) =\exp(-2\gamma t) -1-2 \gamma t $ and $\phi(t) =
\exp(2\gamma t) -1-2 \gamma t $. For short times $ t $ assuming $
t \ll 1/\gamma $ we have $\langle
[\hat{a}_2(t),\hat{a}_2^{\dagger} (t) ] \rangle = 1+ (\epsilon^2
-\kappa^2)t^2 $ and $ \langle [\hat{a}_1
(t),\hat{a}_2^{\dagger}(t) ] \rangle = -2i\epsilon t$. Thus, we
additionally require $ t \ll 1/\epsilon $ and $ t \ll
1/\sqrt{\epsilon^2 -\kappa^2} $. As we usually assume in $
\mathcal{PT} $-symmetric systems that $ \kappa \le \epsilon $, we
are left with the following conditions for applicability of the
model with unidirectional coupling:
\begin{equation}   
 t \ll {\rm min} \left\{ \frac{1}{\gamma}, \frac{1}{\epsilon} \right\}.
\label{59}
\end{equation}

The form of the commutation relations given in Eq.~(\ref{58}) and
the ensuing restricted validity of the model poses the question
about possible corrections of the model using suitable properties
of the reservoir Langevin operator forces. Similarly as it is done
when damping and amplification are introduced into the Heisenberg
equations (the Wigner--Weisskopf model of damping, see
Ref.~\cite{Perina1991}). However, as discussed in detail in
Appendix~\ref{AppC}, this approach is not successful.

Also, in Appendix~\ref{AppD} the properties of the modes are
analyzed in the framework of the Gaussian states, their
nonclassicality depths and logarithmic negativity are determined.
These results point out at specific properties of the two-mode
bosonic system with unidirectional coupling applicable only under
the conditions given in Eq.~(\ref{59}).

These results lead us to the conclusion that this method for
concatenating simpler bosonic systems with QEPs based on
unidirectional coupling to arrive at QEPs with higher-order ED has
limitations. The question how to obtain higher-order inherited
QEPs in bosonic systems without these limitations  {and under
general conditions} is open.

\section{Higher-order hybrid points revealed by field-operator moments}
\label{sec5}

In the last section, we derive general formulas that give us the
orders of EDs and DDs of QHPs that occur in the dynamics of
higher-order FOMs. We note that the structure of higher-order FOM
spaces mapped onto suitable lattices and its influence to system's
properties has been analyzed in detail in Ref.~\cite{Arkhipov2023}
for two- and three-mode bosonic systems.

Let us consider a bosonic system composed of $ n $ modes and
described by an appropriate quadratic non-Hermitian Hamiltonian.
Such a system is described by the $ 2n $ annihilation and creation
operators, and the $ 2n\times 2n $ dynamical matrix $ \bm{M^{(n)}}
$ of its Heisenberg-Langevin equations has $ 2n $ eigenvalues. Let
us fix a position in the system parameter space. After
diagonalization of the  {$ 2n\times 2n $} dynamical matrix $ \bm{M^{(n)}} $, we
identify the eigenvalues $ \Lambda_j $ with
different eigenvectors and count their degeneration numbers $ n_j
$ according to the number of coalescing eigenvectors (from the
original maximal Hilbert space).  {We note that the eigenvalues
$\Lambda_j$ can coincide, leading to diabolical degeneracies. The
corresponding eigenvalue structure is illustrated in
Fig.~\ref{fig4} for reference.}
\begin{figure}  
 \begin{center}
   \begin{tabular}{ccccccc}
\multicolumn{7}{c}{$ \overbrace{ \hspace{60mm} }^{k} $} \cr
\multicolumn{3}{c}{$\underbrace{\Lambda_1 \hspace{3mm} \ldots
\hspace{3mm} \Lambda_1}_{k_1} $} & \ldots \ldots \ldots &
\multicolumn{3}{c}{$ \underbrace{ \Lambda_{\Sigma_\Lambda}
\hspace{3mm} \ldots \hspace{3mm}
\Lambda_{\Sigma_\Lambda}}_{k_{\Sigma_\Lambda}} $} \cr $
\bm{\hat{B}}_{1,1} $ & \ldots & $ \bm{\hat{B}}_{1,1} $ & \ldots
\ldots \ldots & $ \bm{\hat{B}}_{\Sigma_\Lambda,1} $ & \ldots & $
\bm{\hat{B}}_{\Sigma_\Lambda,1} $ \cr \ldots & \ldots &\ldots &
\ldots \ldots \ldots & \ldots & \ldots & \ldots \cr \ldots &
\ldots &\ldots & \ldots \ldots \ldots & \ldots & \ldots & \ldots
\cr $ \bm{\hat{B}}_{1,n_1} $ & \ldots & $ \bm{\hat{B}}_{1,n_1} $ &
\ldots \ldots \ldots &
$\bm{\hat{B}}_{{\Sigma_\Lambda},n_{\Sigma_\Lambda}} $ & \ldots & $
\bm{\hat{B}}_{{\Sigma_\Lambda},n_{\Sigma_\Lambda}} $
\end{tabular}
 \end{center}
 \caption{ {Scheme for constructing general $ k $th-order FOMs for an $ n $-mode bosonic system
  with $ \Sigma_\Lambda $ eigenvalues $ \Lambda_1,\ldots,\Lambda_{\Sigma_\Lambda} $
  having $ n_1, \ldots, n_{\Sigma_\Lambda} $ coalescing eigenvectors, i.e.
  $ \sum_{j=1}^{\Sigma_\Lambda} n_j = 2n $. The field operators
  $ \hat{b}_l $ and $ \hat{b}_l^\dagger $ for $ l=1,\ldots, n $ in
  the diagonalized basis form elements of the field-operator
  vectors $ \bm{\hat{B}_1}, \ldots, \bm{\hat{B}_{\Sigma_\Lambda}} $
  composed of in turn $ n_1, \ldots, n_{\Sigma_\Lambda} $ elements. The $ k $th-order FOMs
  are divided into the groups identified with the vectors $ \bm{k} \equiv [k_1,\ldots,k_{\Sigma_\Lambda}] $,
  $ \sum_{j=1}^{\Sigma_\Lambda} k_j = k $, containing FOMs of the form
  $ \langle \bm{\hat{B}}_{1}^{k_1} \cdots \bm{\hat{B}}_{\Sigma_\Lambda}^{k_{\Sigma_\Lambda}} \rangle $.
  }}
\label{fig4}
\end{figure}
Denoting the number of such eigenvalues by  $ \Sigma_\Lambda $, we
have
\begin{equation}  
 \sum_{j=1}^{\Sigma_\Lambda} n_j = 2n.
\label{60}
\end{equation}
We assign, to any eigenvalue $ \Lambda_j $, an operator vector $
\bm{ \hat{B}_j} $ that encompasses the diagonalized field
operators $ \hat{b}_l $ and $ \hat{b}_l^\dagger $ ($ l=1,\ldots, n
$) that are associated with the coalescing vectors of this
eigenvalue, as shown in Fig.~\ref{fig4}.

Now we consider the dynamics of $ k $th-order FOMs. These $ k
$th-order FOMs can formally be expressed in their general form
using  {the elements of} the above-defined operator vectors $
\bm{ \hat{B}_j} $ as $ \langle \prod_{j=1}^{\Sigma_\Lambda}
\bm{\hat{B}_j}^{k_j} \rangle $, where the nonnegative integers $
k_j $ obey $ \sum_{j=1}^{\Sigma_\Lambda} k_j = k $. The vector $
\bm{k} $ defined as $ [k_1,k_2, \ldots, k_{\Sigma_\Lambda}] $ then
points out at a possible QHP with the complex eigenfrequency $
\Lambda $ given as
 {\begin{equation}   
 \Lambda = \sum_{j=1}^{\Sigma_\Lambda} k_j \Lambda_j .
\label{61}
\end{equation} }
The degrees $ d_{\rm e}^{\bm{k}} $ of ED and $ d_{\rm d}^{\bm{k}}
$ of DD of this QHP are given by the following formulas (compare
the scheme in Fig.~\ref{fig4}):
\begin{eqnarray} 
 d_{\rm e}^{\bm{k}} &=& \prod_{j=1}^{\Sigma_\Lambda} n_j^{k_j} ,
\label{62}\\
 d_{\rm d}^{\bm{k}} &=& \frac{ k! }{ \prod_{j=1}^{\Sigma_\Lambda} k_j!}.
\label{63}
\end{eqnarray}

The number $ N^{(k)}_B $ of different FOMs contributing to DD
observed for the $ k $th-order FOMs (see the right-parts of the
third main columns of Tabs.~II---IV in Appendix~\ref{AppA} and
Tabs.~I---IV in Appendix~B of Ref.~\cite{Thapliyal2025}) is
determined using the combination number  {(permutations with
repetition, $ k $ balls in $ \Sigma_\Lambda $ drawers)} as:
\begin{equation}   
 N^{(k)}_B = \left( \begin{array}{c} \Sigma_\Lambda + k-1 \\ k
 \end{array} \right) = \left( \begin{array}{c} \Sigma_\Lambda + k-1 \\  \Sigma_\Lambda -1
 \end{array} \right).
\label{64}
\end{equation}
The number $ N^{(\bm{k})}_b $ of different $ k $th-order FOMs
written in the operators  $ \hat{b}_l $ and $ \hat{b}_l^\dagger $,
that are the elements of the vectors $ \bm{\hat{B}_j} $, and
belonging to a fixed vector $ \bm{k} $ is expressed as:
 {\begin{equation}   
 N^{(\bm{k})}_b = \prod_{l=1}^{\Sigma_\Lambda} \left( \begin{array}{c} n_l + k_l -1 \\ k_l
 \end{array} \right) = \prod_{l=1}^{\Sigma_\Lambda} \left( \begin{array}{c} n_l + k_l -1 \\ n_l
 -1 \end{array} \right)    .
\label{65}
\end{equation} }

 {We note that a formula similar to Eq.~(\ref{64}) gives the
number $ N^{(\tilde{k})}_b $ of different $ \tilde{k} $th-order
FOMs written in the operators $ \hat{b}_l $ and $
\hat{b}_l^\dagger $ for $ l=1\ldots n $, $ \langle
\hat{b}_1^{\tilde{k}_1}\cdots \hat{b}_l^{\tilde{k}_l} \cdots
\hat{b}_1^{\dagger \tilde{k}_{l+1}}\cdots \hat{b}_l^{\dagger
\tilde{k}_{2l}} \rangle $ for $ \bm{\tilde{k}} = [\tilde{k}_1,
\ldots, \tilde{k}_{2l}] $ and $ \sum_{j=1}^{2l} \tilde{k}_j =
\tilde{k} $ (permutations with repetition, $ \tilde{k} $ balls in
$ 2n $ drawers)}:
\begin{equation}   
 N^{(\tilde{k})}_b = \left( \begin{array}{c} 2n + \tilde{k}-1 \\
 \tilde{k} \end{array} \right) = \left( \begin{array}{c} 2n + \tilde{k}-1 \\ 2n-1
 \end{array} \right) .
\label{66}
\end{equation}
These FOMs are explicitly written in the left-parts of the third
main columns of Tabs.~II---IV in Appendix~\ref{AppA} and
Tabs.~I---IV in Appendix~B of Ref.~\cite{Thapliyal2025}.

To demonstrate the general approach and formulas, we apply them to
the four-mode bosonic system in the circular configuration
analyzed in Sec.~\ref{sec2} in the regime of nonconventional
PT-symmetric dynamics. Its first- and second-order FOMs and the
revealed QHPs with their degeneracies are given in Tab.~\ref{tab2}
in Appendix~\ref{AppA}. The general approach gives us
Tab.~\ref{tab1} that allows to easily derive the content of
Tab.~\ref{tab2} in Appendix~\ref{AppA}: Two lines belonging to $
\bm{\Lambda_j^{\rm i}} = \gamma_+ $ are obtained assuming $ \bm{k}
= (0,0,0,0,1,0) $ and $ (0,0,0,0,0,1) $. For $ \bm{\Lambda_j^{\rm
i}} = 2\gamma_+ $ the entries to Tab.~\ref{tab2} are in turn
generated assuming $ \bm{k} = (0,0,0,0,1,1) $, $ (0,0,0,0,2,0) $,
and $ (0,0,0,0,0,2) $. The orders of the corresponding EDs and DDs
are derived using Eqs.~(\ref{62}) and (\ref{63}), respectively.
\begin{table}[t]    
\begin{center}
\begin{tabular}{|c|c|c|c|}
 \hline
  $ -i \Lambda_j $ & $ n_j $ & $ \bm{\hat{B}_j} $ & $ k_j $ \\
 \hline
 \hline
  $ \gamma_{13} $ & 1 & $ \bm{\hat{B}_1} \equiv [\hat{b}_1 ] $ & $ k_1=0  $ \\
  \hline
  $ \gamma_{13} $ & 1 & $ \bm{\hat{B}_2} \equiv [\hat{b}_1^\dagger ] $ & $ k_2=0  $ \\
  \hline
  $ \gamma_{24} $ & 1 & $ \bm{\hat{B}_3} \equiv [\hat{b}_2 ] $ & $ k_3=0  $ \\
  \hline
  $ \gamma_{24} $ & 1 & $ \bm{\hat{B}_4} \equiv [\hat{b}_2^\dagger ] $ & $ k_4=0  $ \\
  \hline
  $ \gamma_+ $ & 2 & $ \bm{\hat{B}_5} \equiv [\hat{b}_3,\hat{b}_3^\dagger ] $ & $ k_5  $ \\
  \hline
  $ \gamma_+ $ & 2 & $ \bm{\hat{B}_6} \equiv [\hat{b}_4,\hat{b}_4^\dagger ] $ & $ k_6  $ \\
  \hline
  \hline
   $ \sum_{j=1}^{\Sigma_\Lambda} \Lambda_j = \Lambda $  & $ \sum_{j=1}^{\Sigma_\Lambda} n_j =
   2n $ & $ \langle \prod_{j=1}^{\Sigma_\Lambda} \bm{\hat{B}_j}^{k_j}
   \rangle $  & $ \sum_{j=1}^{\Sigma_\Lambda} k_j = k $ \\
  \hline
\end{tabular} 
\end{center}
\vspace{-5mm}
 \caption{Eigenvalues $ \Lambda_j $, their degeneracies $ n_j $, the corresponding field-operator
  vectors $ \bm{\hat{B}_j} $  {defined in the third column of the table}, and their varying powers $ k_j $ for the four-mode bosonic
  system in the circular configuration analyzed in Tab.~\ref{tab2} in Appendix~\ref{AppA}  {under the condition in Eq.~(\ref{8})}. QHPs are uniquely identified by powers of $ k_5 $ and
  $ k_6 $.}
\label{tab1}
\end{table}

These results can be used for detailed discussions of genuine and
induced QHPs and their degeneracies for FOMs of arbitrary orders
and considering different systems with their specific inherited
QHPs observed in the dynamics of field operators governed by the
Heisenberg-Langevin equations. The occurrence of a QHP with $ n^k
$th ED and second-order DD in the dynamics of $ k $th-order FOMs
in a bosonic system with an inherited QHP with $ n $th-order ED
and second-order DD is probably the most valuable result.

\section{Conclusions}
\label{sec6}

Quantum exceptional, diabolical, and hybrid points were analyzed
in simple bosonic systems with unusual properties. The bosonic
systems exhibiting \emph{nonconventional $ \mathcal{PT}
$-symmetric dynamics} characterized by the observation of quantum
exceptional and hybrid points only in certain subspace(s) of the
whole system Liouville space were revealed and their properties
investigated. Nonconventional second-order inherited quantum
exceptional and hybrid points were identified in four-mode bosonic
systems.

Applying the method of concatenating simple bosonic systems via
unidirectional coupling, the conditions for the observation of up
to sixth-order inherited quantum exceptional and hybrid points
were found. However, by analyzing the behavior of the two-mode
bosonic system with unidirectional coupling we have shown that the
bosonic systems with unidirectional coupling are applicable only
for short times in which they ensure the physically consistent
behavior. In short times, more complex bosonic systems with
diverse structures and arbitrary-order exceptional degeneracies
can be built by concatenating simple bosonic systems via
unidirectional coupling of several types. Nevertheless, tailoring
the properties of the Langevin operator stochastic forces in
systems with unidirectional coupling does not enable extending
their applicability to arbitrary times, in contrast to how the
damping and amplification are consistently described.

The operation of two numerical methods for the identification of
quantum exceptional points and their degeneracies, namely (1) the
transformation of a dynamical matrix into its Jordan form and (2)
the introduction of a suitable perturbation $ \delta $ into the
dynamical matrix and its subsequent eigenvalue analysis, was
demonstrated analytically in two-mode bosonic systems.

The exceptional and diabolical degeneracies of inherited quantum
hybrid points were used to derive higher-order degeneracies
observed in the dynamics of higher-order field-operator moments.
The quantum exceptional and hybrid points of second-order
field-operator moments were summarized in tables that evidence a
rich dynamics of the field-operator moments. Numbers of genuine
and induced quantum hybrid points and their exceptional and
diabolical degeneracies were expressed as they depend on the order
of the field-operator moments in the general form using parameters
of the inherited quantum exceptional and hybrid points and their
degeneracies.

The presented analysis considerably broadens the knowledge of
bosonic systems with $ \mathcal{PT} $-symmetry exhibiting
well-accepted physical behavior, though it does not reveal bosonic
systems with higher-order exceptional and hybrid singularities
found under general conditions (e.g. long times).

 {The present analysis, together with the results of
Ref.~\cite{Thapliyal2025}, significantly advances our
understanding of bosonic systems with $\mathcal{PT}$-symmetry and
their spectral singularities. It demonstrates that observing
higher-order exceptional and hybrid singularities in physically
well-behaved (i.e., real) bosonic systems at arbitrary times
remains a challenging task.}

\section{Acknowledgements}

The authors thank Ievgen I. Arkhipov for useful discussions. J.P.
and K.T. acknowledge support by the project OP JAC
CZ.02.01.01/00/22\_008/0004596 of the Ministry of Education,
Youth, and Sports of the Czech Republic. J.P. acknowledges support
by the project No. 25-15775S of the Czech Science Foundation.
A.K.-K., G.Ch., and A.M. were supported by the Polish National
Science Centre (NCN) under the Maestro Grant No.
DEC-2019/34/A/ST2/00081.

\appendix

\section{QEPs and QHPs in first- and second-order FOMs spaces for uni- and bidirectional
coupling} 
\label{AppA}

 {Considering both unidirectional and bidirectional coupling
schemes discussed in the main text, we provide tables listing the
QEPs and QHPs, along with their associated degeneracies, as found
in first- and second-order FOM spaces. Based on the inherited QEPs
and QHPs and their degeneracies observed in the dynamics of
first-order FOMs, we construct the genuine and induced QEPs and
QHPs in the second-order FOM dynamics, following the scheme
outlined in Ref.~\cite{PerinaJr2022a}. This analysis complements
the results presented in Part I of Ref.~\cite{Thapliyal2025} and
relies on the general formulas for spectral degeneracies and their
multiplicities provided in Sec. V. The tables below illustrate the
diversity and richness of spectral degeneracies that can emerge in
simple bosonic models.}

 {The circular four-mode bosonic system described by the
dynamical matrix $ \bm{M^{(4)}_{\rm c}} $ given in Eq.~(\ref{3})
with different damping and/or amplification rates of neighbor
modes [see Eq.~(\ref{5})] exhibits nonconventional
$\mathcal{PT}$-symmetric dynamics. Its QEPs and QHPs found in the
dynamics of the first- and second-order FOMs are given in
Tab.~\ref{tab2}.}
\begin{table*}[ht]    
\begin{center}
 \footnotesize{
\begin{tabular}{|c|c|p{1.3cm}|c|c|c|c|c|c|}
 \hline
  $ { \Lambda^{\rm i}_{j}} $ & $ { \Lambda^{\rm r}_{j}} $ & \multicolumn{2}{c|}{Moments} & Moment & \multicolumn{2}{c|}{Genuine and induced QHPs} &  \multicolumn{2}{c|}{Genuine QHPs}\\
 \cline{6-9}
   &  &  \multicolumn{2}{c|}{} & deg. & Partial & Partial & Partial & Partial \\
  &  &  \multicolumn{2}{c|}{}  &  & QDP x & QDP x & QDP x & QDP x  \\
    &  &  \multicolumn{2}{c|}{}  &  & QEP deg. & QEP deg. & QEP deg. & QEP deg.\\
 \hline
 \hline
  $ \gamma_+ $ & $ \pm \beta $ & $ \langle \hat{b}_3 \rangle $, $ \langle \hat{b}_3^\dagger \rangle $ & $ \langle \bm{\hat{B}_5} \rangle $ & 1 & 1x2 & 2x2  & 1x2 & 2x2\\
 \cline{3-6}   \cline{8-8}
  &  & $ \langle \hat{b}_4 \rangle $, $ \langle \hat{b}_4^\dagger \rangle $ & $ \langle \bm{\hat{B}_6}  \rangle $ &  1 & 1x2 & & 1x2 &  \\
 \hline
  $ 2 \gamma_+ $ & $ \pm 2\beta $ & $ \langle \hat{b}_3 \hat{b}_4 \rangle $, $ \langle \hat{b}_3^\dagger \hat{b}_4^\dagger \rangle $ & $\langle \bm{\hat{B}_5\hat{B}_6} \rangle $ &  2 & 2x4 & 4x4 & 1x4 & 1x4 \\
      & $ \beta - \beta $  &  $ \langle \hat{b}_3^\dagger \hat{b}_4 \rangle $ &  & 2 & & & & + \\
      & $ \beta - \beta $  &  $ \langle \hat{b}_3 \hat{b}_4^\dagger \rangle $ & &  2 &  & & & \\
 \cline{2-6}   \cline{8-8}
  & $ \pm 2\beta $ & $ \langle \hat{b}_3^2 \rangle $, $ \langle \hat{b}_3^{\dagger 2}\rangle $ & $ \langle \bm{\hat{B}_5^2}  \rangle $ &  1 & 1x4 &  & 1x3 & 2x3\\
  & $ \beta - \beta $  & $ \langle \hat{b}_3^\dagger \hat{b}_3 \rangle $ & &  2 & & & & \\
 \cline{2-6}  \cline{8-8}
  & $ \pm 2\beta $ & $ \langle \hat{b}_4^2 \rangle $, $ \langle \hat{b}_4^{\dagger 2}\rangle $ & $ \langle \bm{\hat{B}_6^2}  \rangle $ & 1 & 1x4 &  & 1x3 &  \\
      & $ \beta - \beta $  &  $ \langle \hat{b}_4^\dagger \hat{b}_4 \rangle $ &  &  2 & &  & &\\
  \hline
\end{tabular} }
\end{center}
\vspace{-5mm}
 \caption{Real and imaginary parts of the complex eigenfrequencies $
  {\Lambda_{j}^{\rm r}} - i
  {\Lambda_{j}^{\rm i}} $ of the matrix $ \bm{M^{(4)}_{\rm c}} $, given in Eq.~(\ref{3}) for the four-mode bosonic system in the circular
   configuration with different damping and/or amplification rates of neighbor modes
   valid for the regime of nonconventional $ \mathcal{PT} $ -symmetric dynamics and derived from the equations for the FOMs up to second order. The corresponding moments
   written in the
  `diagonalized' field operators involving the operators $ \hat{b}_3 $, $ \hat{b}_3^\dagger $, $ \hat{b}_4 $, and $ \hat{b}_4^\dagger $
  are given together with their degeneracies (deg.) coming from
  different possible relative positions of the field operators. The DDs of QHPs (partial DDs) derived from
  the indicated FOMs and the EDs of the constituting QEPs are given. Both genuine and induced QEPs and QHPs are considered.
   {The operator vectors $ \bm{\hat{B}_j} $ for
  $ j=5,6 $ are defined in the rows written for $ \bm{ \Lambda^{\rm i}_{j}} = \gamma_+ $
  devoted to the first-order FOMs, i.e. $ \bm{\hat{B}_5} \equiv [\hat{b}_3,\hat{b}_3^\dagger] $,
  $ \bm{\hat{B}_6} \equiv [\hat{b}_4,\hat{b}_4^\dagger] $.
  Symbol $ \bm{\hat{B}_j \hat{B}_k} $, $ j,k=5,6 $, stands for the tensor product
  that gives four terms explicitly written in the rows for $ \bm{ \Lambda^{\rm i}_{j}} = 2\gamma_+
  $; the terms derived from those explicitly written by using the commutation relations are omitted.}}
\label{tab2}
\end{table*}

 {Considering another system giving the QEPs and QHPs with nonconventional
$\mathcal{PT}$-symmetric dynamics --- the
four-mode bosonic system in tetrahedral configuration with the
dynamical matrix $ \bm{M^{(4)}_{\rm t}} $ given in Eq.~(\ref{15})
and equal damping and/or amplification rates of neighbor modes
[see Eq.~(\ref{16})], we reveal the QEPs and QHPs belonging to the
dynamics of the first- and second-order FOMs as summarized in
Tab.~\ref{tab3}.}
\begin{table*}[t]    
\begin{center}
 \footnotesize{
\begin{tabular}{|c|c|p{1.3cm}|c|c|c|c|c|c|}
 \hline
  $ { \Lambda^{\rm i}_{j}} $ & $ { \Lambda^{\rm r}_{j}} $ & \multicolumn{2}{c|}{Moments} & Moment & \multicolumn{2}{c|}{Genuine and induced QHPs} &  \multicolumn{2}{c|}{Genuine QHPs}\\
 \cline{6-9}
   &  &  \multicolumn{2}{c|}{} & deg. & Partial & Partial & Partial & Partial \\
  &  &  \multicolumn{2}{c|}{}  &  & QDP x & QDP x & QDP x & QDP x  \\
    &  &  \multicolumn{2}{c|}{}  &  & QEP deg. & QEP deg. & QEP deg. & QEP deg.\\
 \hline
 \hline
  $ \gamma_+ $ & $ -\zeta\pm\beta$ & $ \langle \hat{b}_3 \rangle $, $ \langle \hat{b}_4^\dagger \rangle $ & $ \langle \bm{\hat{B}_5} \rangle $ & 1 & 1x2 & 1x2  & 1x2 & 1x2\\
 \cline{2-9}
  & $ \zeta\pm\beta$ & $ \langle \hat{b}_4 \rangle $, $ \langle \hat{b}_3^\dagger \rangle $ & $ \langle \bm{\hat{B}_6}  \rangle $ &  1 & 1x2 & 1x2 & 1x2 & 1x2 \\
 \hline
  $ 2 \gamma_+ $ & $ \pm 2\beta $ & $ \langle \hat{b}_3 \hat{b}_4 \rangle $, $ \langle \hat{b}_4^\dagger \hat{b}_3^\dagger \rangle $ & $\langle \bm{\hat{B}_5\hat{B}_6} \rangle $ &  2 & 2x4 & 2x4 & 1x4 & 1x4 \\
      & $ \pm2\zeta$  &  $ \langle \hat{b}_3 \hat{b}_3^\dagger \rangle $, $ \langle \hat{b}_4^\dagger \hat{b}_4 \rangle $ &  & 2 & & & &  \\
 \cline{2-9}
  & $ -2\zeta\pm2\beta$ & $ \langle \hat{b}_3^2 \rangle $, $ \langle \hat{b}_4^{\dagger 2}\rangle $ & $ \langle \bm{\hat{B}_5^2}  \rangle $ &  1 & 1x4 &  1x4 & 1x3 & 1x3\\
  & $ -2\zeta$  & $ \langle \hat{b}_4^\dagger \hat{b}_3 \rangle $ & &  2 & & & & \\
 \cline{2-9}
  & $ 2\zeta\pm2\beta$ & $ \langle \hat{b}_4^2 \rangle $, $ \langle \hat{b}_3^{\dagger 2}\rangle $ & $ \langle \bm{\hat{B}_6^2}  \rangle $ & 1 & 1x4 & 1x4 & 1x3 & 1x3 \\
      & $ 2\zeta$  &  $ \langle \hat{b}_3^\dagger \hat{b}_4 \rangle $ &  &  2 & &  & &\\
  \hline
\end{tabular} }
\end{center}
\vspace{-5mm}
 \caption{Real and imaginary parts of the complex eigenfrequencies $
  {\Lambda_{j}^{\rm r}} - i
  {\Lambda_{j}^{\rm i}} $ of the matrix $ \bm{M^{(4)}_{\rm t}} $, given in Eq.~(\ref{15}),
   for the four-mode bosonic system in the tetrahedral configuration
   valid for the regime of nonconventional $ \mathcal{PT} $-symmetric dynamics
   and derived from the equations for the FOMs up to second order.
    {We have $ \bm{\hat{B}_5} \equiv [\hat{b}_3,\hat{b}_4^\dagger] $,
   $ \bm{\hat{B}_6} \equiv [\hat{b}_4,\hat{b}_3^\dagger] $, and more details are
   given in the caption to Tab.~\ref{tab2}.}}
\label{tab3}
\end{table*}

 {Table~\ref{tab4} presents the QEPs and QHPs observed in the
dynamics of first- and second-order FOMs for the $n$-mode bosonic
models with $n=3,...,6$ and unidirectional coupling. This includes
systems characterized by the dynamical matrices $
\bm{M^{(1+2)}_{\rm u}} $ in Eq.~(\ref{24}) under the condition in
Eq.~(\ref{25}), $ \bm{M^{(2+2)}_{\rm u}} $ in Eq.~(\ref{30}) under
the condition in Eq.~(\ref{32}), $ \bm{M^{(2+3)}_{\rm u}} $ in
Eq.~(\ref{35}) under the condition in Eqs.~(\ref{37}) and
(\ref{40}), and $ \bm{M^{(3+3)}_{\rm u}} $ in Eq.~(\ref{42}) under
the condition in Eq.~(\ref{43}).}
\begin{table*}[ht]    
\begin{center}
\footnotesize{
\begin{tabular}{|p{0.3cm}|p{1.4cm}|c|c|p{1cm}|p{0.8cm}|p{0.8cm}|p{0.8cm}|p{0.8cm}|}
 \hline
  $ { \Lambda^{\rm i}_{j}} $ & $ { \Lambda^{\rm r}_{j}} $ & \multicolumn{2}{p{0.25cm}|}{Moments} & Moment & \multicolumn{2}{p{0.25cm}|}{Genuine and induced QHPs} &  \multicolumn{2}{p{0.25cm}|}{Genuine QHPs}\\
 \cline{6-9}
   &  &  \multicolumn{2}{c|}{} & deg. & Partial & Partial & Partial & Partial \\
  &  &  \multicolumn{2}{c|}{}  &  & QDP x & QDP x & QDP x & QDP x  \\
    &  &  \multicolumn{2}{c|}{}  &  & QEP deg. & QEP deg. & QEP deg. & QEP deg.\\
 \hline
 \hline
  $ \gamma_+ $ & $ \pm \beta_1, \ldots,$ & $ \langle \hat{b}_1 \rangle, \langle \hat{b}_1^\dagger \rangle, \ldots, \langle \hat{b}_{n/2} \rangle, \langle \hat{b}_{n/2}^\dagger \rangle $  & $ \langle \bm{\hat{B}_1} \rangle $ & 1 & 1xn & 2xn  & 1xn & 2xn\\
 \cline{3-6}   \cline{8-8}
  & $\pm \beta_n  $  & $ \langle \hat{b}_{n/2+1} \rangle, \langle \hat{b}_{n/2+1}^\dagger \rangle, \ldots, \langle \hat{b}_n \rangle, \langle \hat{b}_n^\dagger \rangle $     & $ \langle \bm{\hat{B}_2}  \rangle $ &  1 & 1xn & & 1xn &  \\
 \hline
  $ 2 \gamma_+ $ & $ \pm (\beta_k +\beta_l) $ & $ \langle \hat{b}_k \hat{b}_{n/2+l} \rangle $, $ \langle \hat{b}_k^\dagger \hat{b}_{n/2+l}^\dagger \rangle $ & $\langle \bm{\hat{B}_1\hat{B}_2} \rangle $ &  2 & 2x$n^2$ & 4x$n^2$  & 1x$n^2$ & 1x$n^2$ \\
      & $ \beta_l - \beta_k $  &  $ \langle \hat{b}_k^\dagger \hat{b}_{n/2+l} \rangle $ &  & 2 & & & & + \\
      & $ \beta_k - \beta_l $  &  $ \langle \hat{b}_k \hat{b}_{n/2+l}^\dagger \rangle $ & &  2 &  & & & \\
      &  & $ k,l=1,\ldots n/2 $ & & & & & & \\
 \cline{2-6}   \cline{8-8}
  & $ \pm 2\beta_k $ & $ \langle \hat{b}_k^2 \rangle $, $ \langle \hat{b}_k^{\dagger 2}\rangle $ & $ \langle \bm{\hat{B}_1^2}  \rangle $ &  1 & 1x$n^2$ &  & 1x $(n+1)n/2$ & 2x $(n+1)n/2$ \\
      &  & $ k=1,\ldots n/2 $ & & & & & & \\
      & $ \pm(\beta_k + \beta_l) $  &  $ \langle \hat{b}_k \hat{b}_l \rangle $,  $ \langle \hat{b}_k^\dagger \hat{b}_l^\dagger \rangle $  &  & 2 & & & &  \\
      &  & $ k,l=1,\ldots n/2, l < k $  & & & & & & \\
      & $ \beta_l - \beta_k $  &  $ \langle \hat{b}_k^\dagger \hat{b}_l \rangle $ &  & 2 & & & &  \\
      &  & $ k,l=1,\ldots n/2 $ & & & & & & \\
 \cline{2-6}  \cline{8-8}
  & $ \pm 2\beta_k $ & $ \langle \hat{b}_{n/2+k}^2 \rangle $, $ \langle \hat{b}_{n/2+k}^{\dagger 2}\rangle $ & $ \langle \bm{\hat{B}_2^2}  \rangle $ &  1 & 1x$n^2$ &  & 1x $(n+1)n/2$ & \\
      &  & $ k=1,\ldots n/2 $ & & & & & & \\
      & $ \pm(\beta_k + \beta_l) $  &  $ \langle \hat{b}_{n/2+k} \hat{b}_{n/2+l} \rangle $,  $ \langle \hat{b}_{n/2+k}^\dagger \hat{b}_{n/2+l}^\dagger \rangle $  &  & 2 & & & &  \\
      &  & $ k,l=1,\ldots n/2, l<k $  & & & & & & \\
      & $ \beta_l - \beta_k $  &  $ \langle \hat{b}_{n/2+k}^\dagger \hat{b}_{n/2+l} \rangle $ &  & 2 & & & &  \\
      &  & $ k,l=1,\ldots n/2 $ & & & & & & \\
  \hline
\end{tabular} }
\end{center}
\vspace{-5mm}
 \caption{Real and imaginary parts of the complex eigenfrequencies $ {\Lambda_{j}^{\rm r}} - i
  {\Lambda_{j}^{\rm i}} $ of the matrix $ \bm{M^{(n)}_{\rm u}} $ for $ n $-mode bosonic system ($ n>2 $)
   with unidirectional coupling
   having a QHP with $ n $th-order ED and second-order DD derived from the equations
   for the FOMs up to second order. Table is valid for even $ n $, where the
   vector $ \bm{\hat{b}} $ of the diagonalized
   field operators is written as $ \hat{\bm{b}}=[\hat{b}_{1},\hat{b}_{2},\hat{b}_{1}^{\dagger},
  \hat{b}_{2}^{\dagger},\ldots \hat{b}_{n-1},\hat{b}_{n},\hat{b}_{n-1}^{\dagger},\hat{b}_{n}^{\dagger}] $.
  The vector $ \bm{\hat{b}} $ attains the form  $
  \hat{\bm{b}}=[\hat{b}_{1},\hat{b}_{1}^\dagger,\ldots,\hat{b}_{m},\hat{b}_{m}^\dagger,\ldots,
  \hat{b}_{m+1},\hat{b}_{m+2},\hat{b}_{m+1}^{\dagger},\hat{b}_{m+2}^{\dagger},\ldots \hat{b}_{n-1},\hat{b}_{n},
  \hat{b}_{n-1}^{\dagger},\hat{b}_{n}^{\dagger}] $ if there exist $ m $
  un-paired eigenvalues in the subsystems that compose the analyzed bosonic system
  [see, e.g., Eqs.~(23), (47), and (54) in Ref.~\cite{Thapliyal2025}]. In such cases, the columns entitled
  Moments have to be modified accordingly, but
  all other columns remain valid.
   {We have $ \bm{\hat{B}_1} \equiv [\hat{b}_1, \hat{b}_1^\dagger, \ldots, \hat{b}_{n/2}, \hat{b}_{n/2}^\dagger] $,
   $ \bm{\hat{B}_2} \equiv [\hat{b}_{n/2+1}, \hat{b}_{n/2+1}^\dagger, \ldots, \hat{b}_n,
  \hat{b}_n^\dagger] $, and more details are
   given in the caption to Tab.~\ref{tab2}. For $ m > 0 $
   un-paired eigenvalues in the subsystems, additional operators
   $ \bm{\hat{B}_3}, \ldots, \bm{\hat{B}_{2m+2}} $ have to be
   introduced and the table has to be extended.} }
 \label{tab4}
\end{table*}

 {In Tabs.~\ref{tab2}---\ref{tab4}, we can see the most typical
feature of QEPs and QHPs in higher-order FOMs: The second ($ n
$-th) -order ED observed in first-order FOMs gives raise in
Tabs.~\ref{tab2} and \ref{tab3} (\ref{tab4}) the fourth ($n^2$-th)
-order ED in second-order FOMs. These are examples of the general
rule that assigns $ n^k $ ED in $ k $-th-order FOMs provided that
there occurs $n$-th-order ED in first-order FOMs
\cite{PerinaJr2022a}.}

\section{Numerical identification of exceptional points and their
degeneracies} 
\label{AppB}

The performance of two different numerical approaches to reveal
exceptional points and their orders of exceptional degeneracies is
demonstrated analytically considering the simplest two-mode
bosonic systems, one with the usual bidirectional coupling, the
other with unidirectional coupling.

\subsection{Jordan canonical form of a general matrix}

The Jordan form $ \bm{J_M} $ of a matrix $ \bm{M} $ contains
nonzero elements on the diagonal and also the nearest upper
diagonal. It has the dimension of the matrix  $ \bm{M} $. Some
elements at the nearest upper diagonal equal one if the matrix $
\bm{M} $ is non-diagonalizable. The neighbor elements equal to one
form groups. The number of elements in a given group gives the
order of ED (equal to the number of elements + 1) related to the
corresponding eigenvalue.

For example and considering the dynamical matrix $ \bm{M^{(2)}} $,
given in Eq.~(\ref{29}), belonging to the two-mode bosonic system
with usual coupling under the condition ${\beta}=0$, i.e., where a
second-order QEP occurs, the Jordan form $ \bm{J_{M^{(2)}}} $ and
the corresponding similarity transformation $ \bm{S_{M^{(2)}}} $
such that
\begin{equation} 
 \bm{M^{(2)} = S_{M^{(2)}} J_{M^{(2)}} S^{-1}_{M^{(2)}} }
\label{B1}
\end{equation}
take the form:
\begin{equation} 
  \bm{J_{M^{(2)}}} = \left[ \begin{array}{cc}
     -i\gamma_{+} & 1 \\
     0 & -i\gamma_{+}  \end{array} \right], \hspace{2mm}
  \bm{S_{M^{(2)}}}=\left[ \begin{array}{cc}
   i & -1/\gamma_{-} \\
    1 & 0
  \end{array} \right].
\label{B2}
\end{equation}
In Eq.~(\ref{B2}), the element (1,2) of the matrix $
\bm{J_{M^{(2)}}} $, equal to 1, identifies a second-order QEP.

Similarly, when analyzing the two-mode bosonic system with
unidirectional coupling and the dynamical matrix
$\bm{M^{(1+1)}_{\rm u}} $, written in Eq.~(\ref{20}), under the
condition $ \gamma_{1}=\gamma_{2} $ guaranteeing the existence of
a second-order QEP, we arrive at:
\begin{equation} 
  \bm{J_{M^{(1+1)}}} =\left[ \begin{array}{cc}
   -i\gamma_{2}/2 & 1 \\
    0 & -i\gamma_{2}/2
  \end{array} \right], \hspace{2mm}
  \bm{S_{M^{(1+1)}}}=\left[ \begin{array}{cc}
   0 & 1/\xi \\
    1 & 0
  \end{array} \right].
\label{B3}
\end{equation}

\subsection{Perturbation of a dynamical matrix}

The second approach is based upon introducing a suitable small
perturbation on a dynamical matrix $ \bm{M} $ that can remove both
EDs and DDs. In this approach, the eigenvalues of the perturbed
dynamical matrix $ \bm{M_\delta} $ are determined and the
degeneracies are removed as the degenerated eigenvalues split and
then gradually diverge with the increasing perturbation.
Perturbation can also be used for characterizing ED
\cite{Znojil2019,Wiersig2022a,Wiersig2022b}.

Identifying EDs, we demonstrate different kinds of the influence
of perturbation $ \delta $ on the eigenvalues of the matrix $
\bm{M} $ considering the two-mode systems with the usual and
unidirectional couplings and different positions of the
perturbation $ \delta $ inside the matrix $ \bm{M} $.
Specifically, we demonstrate splitting of the eigenvalues in their
real and/or imaginary parts and splitting proportional to $
\sqrt{\delta} $ and $ \delta $ when a QEP with second-order ED is
disturbed. We quantify the strength of the perturbation by the
overlap $ F $ of the normalized eigenvectors $ \bm{y_1} $ and $
\bm{y_2} $ that become gradually distinguishable as the
perturbation $ \delta $ increases
 {\begin{equation}  
 F = \frac{ | \langle \bm{y_2}|\bm{y_1} \rangle | }{ \sqrt{ \langle \bm{y_1}|\bm{y_1} \rangle
   \langle \bm{y_2}|\bm{y_2} \rangle } }
\label{B4}
\end{equation} }
and symbol $ \langle |\rangle $ stands for the scalar product of
complex vectors.

\clearpage

\begin{enumerate}
 \item A two-mode system with bidirectional coupling and $\beta=0$ described
  by the perturbed dynamical matrix
  \begin{equation}  
    \bm{M^{(2)}_{\delta,1} } = \bm{M^{(2)}} + \left[  \begin{array}{cc}
     \delta  & 0 \\
     0 & 0  \end{array} \right].
  \label{B5}
  \end{equation}
  The eigenvalues and the corresponding eigenvectors take, respectively, the
  form:
  \begin{eqnarray}  
   \lambda_{1,2}^{M^{(2)}_{\delta,1} } = -i \gamma_{+}+\frac{\delta}{2} \mp \sqrt{\delta\left(\frac{\delta}{4}-i \gamma_{-} \right)}
   \label{B6}   
   \end{eqnarray}
   and
   \begin{eqnarray}   
    \bm{y_{1,2}^{M^{(2)}_{\delta,1}} } = \left[i-\frac{\delta
     \mp \sqrt{\delta\left(\delta-4i \gamma_{-} \right)}}{2\gamma_{-}} , 1 \right]^T.
   \label{B7}   
   \end{eqnarray}

\item  A two-mode system with bidirectional coupling and $\beta=0$
  described by the perturbed dynamical matrix
  \begin{equation}  
    \bm{M^{(2)}_{\delta,2} } = \bm{M^{(2)}} + \left[  \begin{array}{cc}
     0 & -\delta  \\
     0 & 0  \end{array} \right].
  \label{B8}
  \end{equation}
  The eigenvalues and the corresponding eigenvectors are derived,
  respectively, as follows:
  \begin{eqnarray}  
   {\lambda_{1,2}^{M^{(2)}_{\delta,2}} } = -i \gamma_{+} \mp \sqrt{\delta \gamma_{-}}
  \label{B9}   
  \end{eqnarray}
  and
  \begin{eqnarray}   
    \bm{y_{1,2}^{M^{(2)}_{\delta,2}} }& =&\left[i\pm \sqrt{\frac{\delta}
    {\gamma_{-}}} , 1 \right]^T.
  \label{B10}   
  \end{eqnarray}

\item  A two-mode system with unidirectional coupling and
  $\gamma_{1}=\gamma_{2}$ described by the perturbed dynamical matrix
  \begin{equation}  
    \bm{M^{(1+1)}_{{\rm u},\delta} } = \bm{M^{(1+1)}_{\rm u}} + \left[  \begin{array}{cc}
     \delta & 0  \\
     0 & 0  \end{array} \right].
  \label{B11}
  \end{equation}
  The eigenvalues and the corresponding eigenvectors are obtained,
  respectively, in the form:
  \begin{eqnarray}  
   {\lambda_{1}^{M^{(1+1)}_{{\rm u},\delta}} }& =&-i \gamma_{2}/2,\nonumber \\
    {\lambda_{2}^{M^{(1+1)}_{{\rm u},\delta}} }& =&-i \gamma_{2}/2+\delta
  \label{B12}   
  \end{eqnarray}
  and
  \begin{eqnarray}   
   \bm{y_{1}^{M^{(1+1)}_{{\rm u},\delta}} }& =&\left[0 , 1 \right]^T, \nonumber \\
   \bm{y_{2}^{M^{(1+1)}_{{\rm u},\delta}} }& =&\left[\frac{\delta}{\xi} , 1 \right]^T.
  \label{B13}   
  \end{eqnarray}
\end{enumerate}
\begin{figure}  
 \begin{centering}
  (a) \includegraphics[width=0.28\hsize]{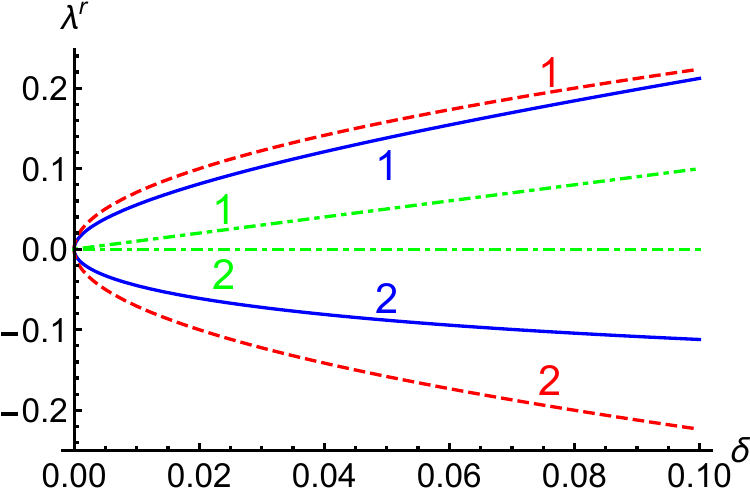}
  (b) \includegraphics[width=0.28\hsize]{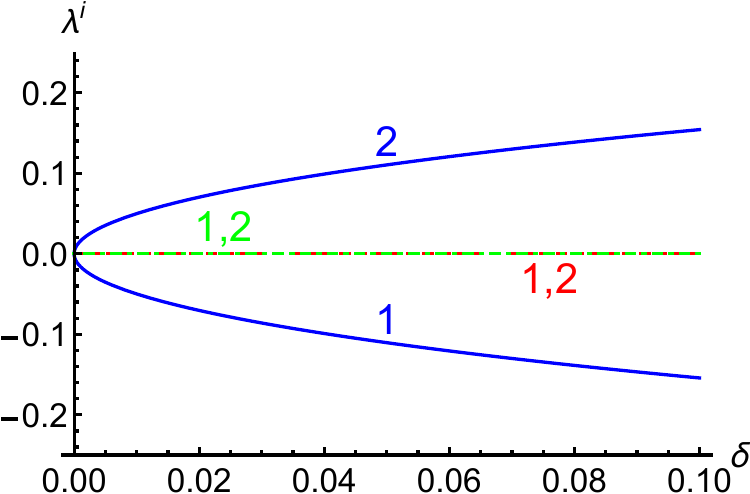}
  (c) \includegraphics[width=0.28\hsize]{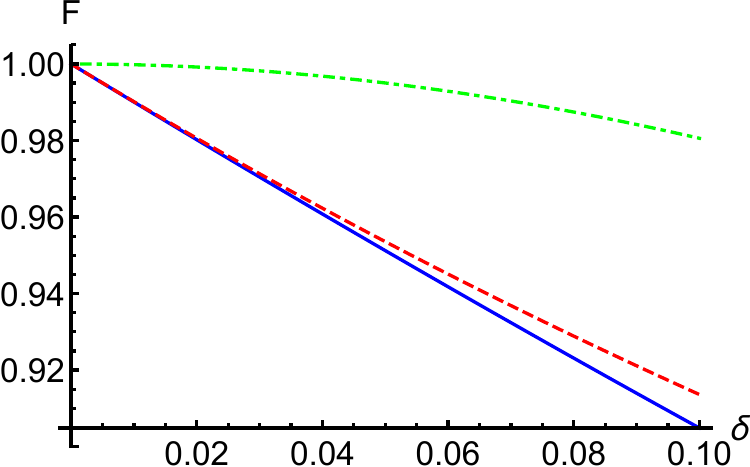}

 \end{centering}
 \caption{(a) Real $ \lambda^{\rm r} $ and (b) imaginary $ \lambda^{\rm i} $ parts of
  eigenvalues $ \lambda_{1,2}$ of the dynamical
  matrices $ \bm{M^{(2)}_{\delta,1} } $ ($\gamma_{+}=0$, $\gamma_{-}=0.5$, blue solid curves), $ \bm{M^{(2)}_{\delta,2}}$
  ($\gamma_{+}=0$, $\gamma_{-}=0.5$, red dashed curves),
  and $ \bm{M^{(1+1)}_{{\rm u},\delta} } $ ($\gamma_{2}=0$, green dot-dashed curves) given
  in turn in Eqs.~(\ref{B5}), (\ref{B8}), and (\ref{B11})
  as they depend on perturbation parameter $ \delta $. In (c) the
  overlap $ F $ of the corresponding eigenvectors is plotted. In (a) and (b), the numbers denote the curves of the corresponding eigenvalues.}
\label{fig5}
\end{figure}

The real and imaginary parts of the eigenvalues $ \lambda_{j}$, $
j=1,2 $, from Eqs.~(\ref{B6}), (\ref{B9}), and (\ref{B12}) are
plotted in Figs.~\ref{fig5}(a,b) as they depend on the
perturbation $ \delta $. The perturbation $ \delta $ in general
disturbs more strongly the system with the usual coupling, as it
is apparent both from the graphs of the eigenvalues and the
overlap $ F $ of eigenvectors shown in Fig.~\ref{fig5}(c).

In the above-discussed cases, the second-order DD present in both
two-mode systems was not modified by the perturbation $ \delta $
because of the structure of these systems. However, suitable
positioning of the perturbation $ \delta $ inside a dynamical
matrix $ \bm{M} $ may also result in revealing the DDs. As the DD
is embedded in the $ 2\times 2 $ matrix $ \bm{\xi} $ given in
Eq.~(\ref{4}), the perturbation $ \delta $ has to affect this
matrix. We consider two kinds of perturbation in the two-mode
system with the dynamical matrix $ \bm{M^{(2)}} $: The first one
splits all eigenvalues $ \Lambda_j $, $ j=1,\ldots,4 $, in their
real parts, whereas the second one distinguishes two eigenvalues
in their real parts and the remaining two eigenvalues in their
imaginary parts. We note that the perturbation $ \delta $
primarily modifies the eigenvalues of the matrix $ \bm{\xi} $,
which removes the DD. Secondarily, as the eigenvalues of $
\bm{\xi} $ determined for nonzero $ \delta $ differ from those
valid for  $ \delta=0 $, the conditions for having a QEP of the
matrix $ \bm{M} $ change and the original setting of the system
parameters for a QEP is lost and so also the corresponding ED is
lost.

\begin{figure}  
 \begin{centering}
  (a) \includegraphics[width=0.45\hsize]{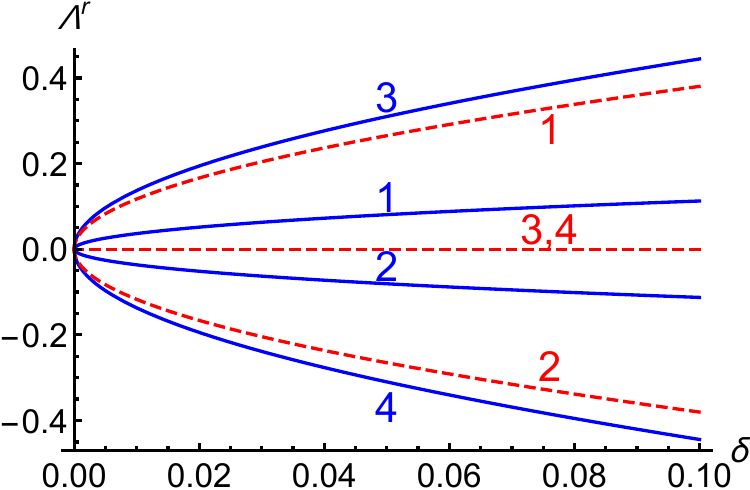}
  (b) \includegraphics[width=0.45\hsize]{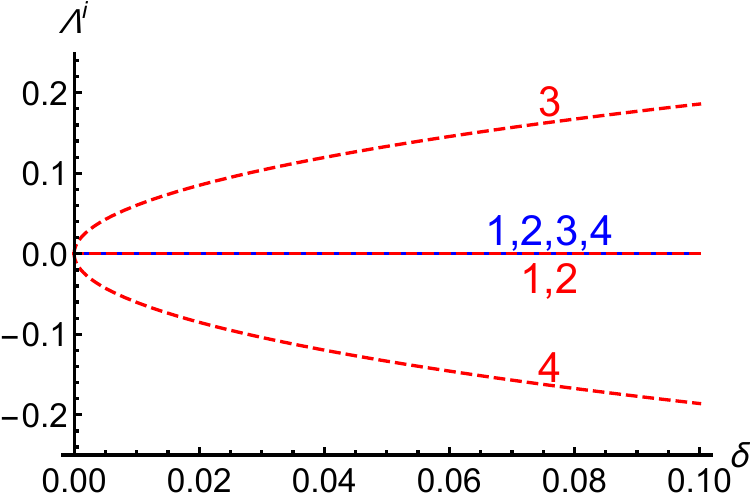}
 \end{centering}
 \caption{(a) Real $ \Lambda^{\rm r} $ and (b) imaginary $ \Lambda^{\rm i} $ parts of the
  eigenvalues $ \Lambda_{1,\ldots,4}$ of the dynamical
  matrix $ \bm{M^{(2)}} $ involving $ \bm{\xi}_{\delta,1} $ (blue solid curves) and $ \bm{\xi}_{\delta,2}$
  (red dashed curves) given in Eqs.~(\ref{B14}) and (\ref{B17}), respectively, as they depend on perturbation parameter $ \delta $;
  $\gamma_{+}=0$, $\beta=0$, $\kappa/\epsilon=1/2$. In (a) and (b), the numbers denote the curves of the corresponding eigenvalues.}
 \label{fig6}
\end{figure}

\begin{enumerate}
 \item A two-mode system with bidirectional coupling and $\beta=0$ described
  by the dynamical matrix $ \bm{M^{(2)}} $ where
  \begin{equation}  
    \bm{\xi}_{\delta,1} = \bm{\xi} + \left[ \begin{array}{cc}
      \delta & 0 \\
      0 & 0
     \end{array} \right].
  \label{B14}
  \end{equation}
  The eigenvalues and the corresponding eigenvectors of matrix $ \bm{\xi}_{\delta,1} $ are
  written, respectively, as:
  \begin{eqnarray} 
    \lambda^{\xi_{\delta,1}}_{1,2}& =&\frac{\delta}{2}\mp \sqrt{\zeta^2+ \delta\left(\epsilon  +\frac{\delta}{4}\right)}
  \label{B15}  
  \end{eqnarray}
   and
  \begin{eqnarray}  
    \bm{y^{\xi_{\delta,1}}_{1,2}}& =&\left[-\frac{\epsilon + \lambda^{\xi_{\delta,1}}_{1,2}}{\kappa} , 1 \right]^T.
  \label{B16} 
  \end{eqnarray}

 \item A two-mode system with bidirectinal coupling and $\beta=0$ described
  by the dynamical matrix $ \bm{M^{(2)}} $ where
  \begin{equation}  
    \bm{\xi}_{\delta,2} = \bm{\xi} + \left[ \begin{array}{cc}
      \delta & \delta \\
      0 & 0
     \end{array} \right].
  \label{B17}
  \end{equation}
  The eigenvalues and the corresponding eigenvectors of matrix $ \bm{\xi}_{\delta,2} $ are
  obtained, respectively, as:
  \begin{eqnarray} 
    \lambda^{\xi_{\delta,2}}_{1,2}& =&\frac{\delta}{2}\mp \sqrt{\zeta^2+ \delta\left(\epsilon -\kappa +\frac{\delta}{4}\right)}
  \label{B18}  
  \end{eqnarray}
  and
  \begin{eqnarray}  
   \bm{y^{\xi_{\delta,2}}_{1,2}}& =&\left[-\frac{\epsilon + \lambda^{\xi_{\delta,2}+\delta}_{1,2}}{\kappa} , 1 \right]^T.
  \label{B19} 
  \end{eqnarray}
\end{enumerate}
The looked for eigenvalues and eigenvectors are then reached using
the eigenvalues and eigenvectors of the $ 2\times 2 $ matrix
$\bm{M^{(2)}}$ written in Eq.~(\ref{29}):
\begin{eqnarray}  
 {\lambda_{1,2}^{M^{(2)}} }& =&-i \gamma_{+} \mp \beta
\label{B20}   
\end{eqnarray}
and
\begin{eqnarray}   
 \bm{y_{1,2}^{M^{(2)}} }& =&\left[-\frac{i
 \gamma_{-}\pm \beta}{{\xi}} , 1 \right]^T,
\label{B21}   
\end{eqnarray}
where $ 4\gamma_{\pm} = \gamma_{1} \pm \gamma_{2}$ and $
\beta^2={{\xi}^2-\gamma_{-}^2}$. The real and imaginary parts of
the eigenvalues $ \Lambda_j $, $ j=1,\ldots,4 $, of the $ 4\times
4 $ matrix $\bm{M^{(2)}}$ are plotted in Fig.~\ref{fig6}.

Finally, we mention a specific case of the perturbation $ \delta $
that removes DD but keeps one from the two originally
diabolically-degenerated QEPs present in the system. The dynamical
matrix $ \bm{M^{(2)}} $ of the two-mode bosonic system with
bidirectional coupling is perturbed in the following way:
\begin{eqnarray}  
 \bm{\tilde{\gamma}_1}^{\delta,3} &=& \bm{\tilde{\gamma}_1} + \left[ \begin{array}{cc}
      \delta & 0 \\
      0 & 0
     \end{array} \right].
\label{B22}
\end{eqnarray}
The eigenvalues $\Lambda^{M^{(2)}_{\delta,3}} $ of the dynamical
matrix in Eq.~(\ref{B22}) are derived in the form:
\begin{eqnarray}   
 \Lambda_{1,2}^{M^{(2)}_{\delta,3}} &=& -i \gamma_{+}\mp\beta, \nonumber \\
 \Lambda_{3,4}^{M^{(2)}_{\delta,3}} &=& -i \gamma_{+}+\frac{\delta}{2}\mp
  \sqrt{\beta^2+ \delta\left(\frac{\delta}{4}-i \gamma_{-} \right) }.
\label{B23}   
\end{eqnarray}
The corresponding eigenvectors are expressed as follows:
\begin{eqnarray} 
 \bm{Y_{1,2}^{M^{(2)}_{\delta,3}}}&= &
  \left[ \begin{array}{cccccc}
   0, & \frac{i \gamma_{-}\pm\beta}{\epsilon}, & -\frac{\kappa}{\epsilon}, & 1
   \end{array}\right]^T , \nonumber \\
 \bm{Y_{3,4}^{M^{(2)}_{\delta,3}}} &= &
  \left[ \begin{array}{cccccc}
 \frac{i \gamma_{-}- \Lambda_{3,4}^{M^{(2)}_{\delta,3}}}{\kappa}, & 0, & -\frac{\epsilon}{\kappa}, & 1
   \end{array}\right]^T .
\label{B24}   
\end{eqnarray}
According to Eqs.~(\ref{B23}) and (\ref{B24}), we observe a single
QEP with second-order ED for $\beta=0$ and $ \delta \neq 0 $. This
contrasts with the observation of a QEP with second-order ED and
also second-order DD for $\beta=0$ and $ \delta = 0 $.

\section{Two-mode bosonic system with unidirectional coupling and relevant reservoir
properties}  
\label{AppC}

Considering the approach outlined in Ref.~\cite{PerinaJr2023}, we
construct the matrix  $ \langle \hat{\bm{L^{\rm u}}}(t)
\bm{\hat{L}^{{\rm u}\dagger}}(t) \rangle $ of the stochastic
Langevin operator forces such that the bosonic commutation
relations of the field operators are obeyed for an arbitrary time
$ t $.

First, we note that the additional, unwanted, terms in the
commutation relations in Eq.~(\ref{58}) disappear when we replace
the correlation matrix $ \langle
\hat{\bm{F}}(t)\hat{\bm{F}}^{\dagger}(t) \rangle $ of the
fluctuating operator forces $ \hat{\bm{F}} $ given in
Eq.~(\ref{56}) by the matrix $ \langle \hat{\bm{F}^{\rm
u}}(t)\hat{\bm{F}}^{{\rm u}\dagger }(t) \rangle $ with the
constituting submatrices:
\begin{eqnarray}   
 \bm{F_{1}^{\rm u}}(t) &=& \bm{F_{1}}(t), \nonumber \\
 \bm{F_{2}^{\rm u}}(t) &=& \bm{F_{2}}(t) + \frac{\phi(t)}{2\gamma^2}
   \left[ \begin{array}{cc}  -\epsilon^2 & 0 \\
   0  & \kappa^2 \end{array}  \right] , \nonumber \\
 \bm{F_{12}^{\rm u}} (t) &=& \bm{F_{12}}(t) + \frac{i \psi(t)}{2\gamma}
  \left[ \begin{array}{cc} \epsilon & 0 \\
   0  & 0  \end{array} \right],
\label{C1}
\end{eqnarray}
where the functions $ \phi $ and $ \psi $ are defined below
Eq.~(\ref{58}).

Inverting Eq.~(\ref{53}) (for details, see Eq.~(18) in
Ref.~\cite{PerinaJr2023}) we obtain the correlation matrix $
\langle\hat{\bm{L^{\rm u}}}(t) \hat{\bm{L}^{{\rm
u}\dagger}}(t')\rangle $ of the Langevin operator forces as
follows:
\begin{eqnarray}  
 \langle\hat{\bm{L^{\rm u}}}(t) \hat{\bm{L}^{{\rm u}\dagger }}(t')\rangle &=& \delta(t-t')
 \left[ \begin{array}{cccc}
   2\gamma & 0 & -i \epsilon l_1(t) & 0 \\
  0 & 0 & 0 & 0 \\
  i \epsilon l_1(t)  & 0 & -\frac{\epsilon^2\mu(t)^2 l_2(t) }{\gamma} & \frac{\epsilon\kappa l_1(t)l_2(t) }{2\gamma} \\
  0 & 0 & \frac{\epsilon\kappa l_1(t)l_2(t)}{2\gamma} & 2\gamma - \frac{\kappa^2 l_2(t)}{\gamma}
 \end{array} \right],
\label{C2}
\end{eqnarray}
where $ l_{1,2} (t) = \exp(-2\gamma t) \pm1$; $\mu(t)=\exp(-\gamma t)$. 

However, the correlation matrix $ \langle\hat{\bm{L^{\rm u}}}(t)
\hat{\bm{L}^{{\rm u}\dagger}}(t')\rangle $  does not represent a
physical reservoir, as it has a negative eigenvalue. This can be
even analytically confirmed considering $ t=0 $. In this case, the
eigenvalues $ \nu $ are obtained as:
\begin{equation}  
 \nu_{1,2,3,4} =  0,\; 2\gamma, \; \gamma \pm
 \sqrt{\gamma^2+4\epsilon^2}.
\label{C3}
\end{equation}
This means in general that there does not exist a physical
reservoir with properties such that the model with unidirectional
coupling could be applied for an arbitrary time $ t $.

\section{Statistical properties of a two-mode bosonic system with unidirectional
coupling}
\label{AppD}

We consider modes 1 and 2 being in their initial coherent states $
\alpha_1 $ and $ \alpha_2 $, respectively. The modes evolution
described in Eq.~(\ref{54}) maintains the state Gaussian form
described by the following normal characteristic function
\cite{Perina1991,PerinaJr2000,PerinaJr2019c}:
\begin{eqnarray}  
 C_{\cal N}(\beta_1,\beta_2;t) &=& \exp\left[ -B_{2}(t)|\beta_{2}|^2 +\{C_{2}(t)\beta_{2}^{*2} + {\rm c.c.} \}/2 \right] \nonumber \\
 & & \hspace{-15mm} \times \exp[ \alpha_1^*(t) \beta_1 + \alpha_2^*(t) \beta_2 -D(t)\beta_1^*\beta_2^*
   + {\rm c.c.}],
\label{D1}
\end{eqnarray}
where symbol c.c. replaces the complex conjugated terms. The
functions $ B_2(t) $, $ C_2(t) $, and $ D(t) $ are given as:
\begin{eqnarray}   
 B_{2}(t) &=& \langle \delta \hat{a}_{2}^{\dagger}(t) \delta \hat{a}_{2}(t)\rangle
  = \frac{\kappa^2\psi(t)}{2\gamma^2}- \frac{ l_1(t)}{ \mu^{2}(t)} , \nonumber \\
C_{2}(t) &=& \langle [\delta \hat{a}_{2}(t) ]^2\rangle
  =  -\frac{\epsilon\kappa \psi(t)}{2\gamma^2}, \nonumber \\
 D(t) &=&  \langle \delta \hat{a}_{1} (t) \delta \hat{a}_{2}(t)\rangle
  = -i \kappa t;
\label{D2}
\end{eqnarray}
and $ \delta \hat{x} = \hat{x} - \langle \hat{x}\rangle $ for an
arbitrary operator $ \hat{x} $. The functions $ \phi $, $ \psi $,
$ \mu $, and $ l_1 $ are defined below Eqs.~(\ref{58}) and
(\ref{C2}), while the complex modes amplitudes $ \alpha_1(t) $ and
$ \alpha_2(t) $ are derived from their initial values along the
relations
\begin{eqnarray}  
 \left[ \begin{array}{c} \alpha_1(t) \\ \alpha_2(t) \end{array} \right] & = &
  \bm{U}(t) \left[ \begin{array}{c} \alpha_1(0) \\ \alpha_2(0) \end{array} \right]
   + \bm{V}(t) \left[ \begin{array}{c} \alpha_1^*(0) \\ \alpha_2^*(0) \end{array}
   \right],
\label{D3}
\end{eqnarray}
where the matrices $ \bm{U}(t) $ and $ \bm{V}(t) $ are given in
Eq.~(\ref{55}).

Using the formulas in Eq.~(\ref{D2}), the nonclassicality depth $
\tau_2 $ of mode 2 \cite{Lee1991}, determined as
\begin{eqnarray}  
  \tau_{2}(t) &=& {\rm max} \bigl\{0, |C_2(t)|-B_{2}(t)\bigr\},
\label{D4}
\end{eqnarray}
is vanishing, which means the classical behavior of the marginal
field in mode 2. Also the marginal field in mode 1 stays
classical, as we have $ \tau_1 = 0 $.

We quantify the entanglement between the fields in modes 1 and 2
by the logarithmic negativity $ E_N $ \cite{Hill1997} determined
from the simplectic eigenvalues $ \nu_{\pm} $ of the
partially-transposed symmetrically-ordered covariance matrix $
\bm{\sigma}^{\rm PT} $ \cite{Adesso2007}, as given by:
\begin{eqnarray} 
 \bm{\sigma}^{\rm PT}(t)  &=& \left[\begin{array}{cc}
  \bm{1} & \bm{\sigma}_{12}^{\rm PT}(t) \\
  \left[\bm{\sigma}_{12}^{\rm PT}(t)\right]^{\rm T} &
  \bm{\sigma}_{2}^{\rm PT}(t)
   \end{array}\right],
\label{D5} \\
  \bm{\sigma}_2^{\rm PT}(t)  &=& \left[\begin{array}{cc}
   1+2B_2(t)+2 C_2(t) & 0\\
   0 & 1+2B_2(t)-2 C_2(t)  \end{array}\right],\nonumber \\
  \bm{\sigma}_{12}^{\rm PT}(t) &=& 2\left[\begin{array}{cc}
  0 & - {\rm Im}\{D(t)\} \\
   {\rm Im}\{D(t)\} & 0 \end{array}\right] ,
\label{D6}
\end{eqnarray}
where $\bm{1}$ stands for the two-dimensional identity matrix.
Determining two invariants $\delta = 1+ {\rm
Det}\{\bm{\sigma}_{2}^{\rm PT} \}+2 {\rm
Det}\{\bm{\sigma}_{12}^{\rm PT} \} $ and $\Delta = {\rm
Det}\{\bm{\sigma}^{\rm PT} \}$ of the matrix $ \bm{\sigma}^{\rm
PT} $, the simplectic eigenvalues $ \nu_{\pm} $ are expressed as
\cite{Adesso2007}:
\begin{eqnarray}  
 2\nu_{\pm}^{2} &=& \delta \pm \sqrt{\delta^2-4\Delta}.
\label{D7}
\end{eqnarray}
The logarithmic negativity $ E_N $ is then determined along the
formula
\begin{equation} 
 E_{N} = {\rm max} \bigl\{0, -\ln(\nu_{-})\bigr\}.
\label{D8}
\end{equation}

As the model is applicable only for short times $ t $, fulfilling
the conditions in Eq.~(\ref{59}), we derive the logarithmic
negativity $ E_N $ using the Taylor expansion in $ t $ to first
order:
\begin{eqnarray}  
 E_{N}(t) &=& -\ln\left[1+2\gamma t \left(1-\sqrt{1+ (\kappa/\gamma)^{2} }\right)
 \right].
\label{D9}
\end{eqnarray}
The formula~(\ref{D9}) predicts nonzero negativity $ E_N $ for
short times $ t $ despite the fact that mode 1 remains in a
coherent state. If $ \alpha_1(0) = (0,0) $ then $ \alpha_1(t) =
(0,0) $ and mode 1 remains in the vacuum state. However, this
state cannot be entangled with mode 2, as the formula~(\ref{D9})
for negativity $ E_N $ predicts. This is another manifestation of
the limited applicability of the studied model with unidirectional
propagation.

\bibliographystyle{quantum}
\bibliography{perina}

\end{document}